\documentclass[12pt,preprint]{aastex}
\begin{document}
\title{Light and Motion in the Local Volume}
\author{Alan B. Whiting}
\affil{Cerro Tololo Inter-American Observatory}
\affil{Casilla 603, La Serena, Chile}
\email{awhiting@noao.edu}

\begin{abstract}
Using high-quality data on 149 galaxies within 10 megaparsecs (Mpc),
I find no correlation between luminosity and peculiar velocity at
all.  There is no unequivocal sign on scales of 1-2 Mpc of the
expected gravitational effect of the brightest galaxies, in particular
infall toward groups; or of infall toward the Supergalactic Plane
on any scale.  Either
dark matter is not distributed in the same way as luminous matter
in this region, or peculiar velocities are not due to fluctuations
in mass.  The sensitivity of peculiar velocity studies to
the background model is highlighted.
\end{abstract}

\section{Introduction}

The first subject of cosmology is the overall motion of the universe.
In terms of the current understanding, this means figuring out
whether it is open or closed, flat or curved, dense or rarified; and 
determining a few numbers ($H_0, \Omega$) which describe it.  In
the past few years this task seems to have been carried almost
to completion.

The next task is to study the details of the universe, that is,
the departures from smoothness and overall motion.  These take the form of
the visible features (galaxies, as well as collections of them in groups,
clusters, superclusters, and so forth) and their peculiar motions.
On the largest scales this may be approached by using perturbations
of the background cosmology, where linearity makes the calculations
easier (or possible at all).  Here, the picture of structure
growing by gravitational instability from small initial fluctuations  
in a smooth distribution of dark and luminous matter seems to work
well (see \citet{CSW00} for a review; more recent work includes
\citet{BZP00} and \citet{NDB01}).  There are disagreements in detail,
for instance over the size of the bias parameter between dark and
luminous matter, but the overall picture shows no great 
difficulties.

The problem of the formation of individual galaxies is much complicated
by the inclusion of radiation and gas effects, and it is beyond
the scope of this paper even to outline the subfield.  For our 
purposes it suffices to note that dark matter and luminous matter are
not distributed the same way on galactic scales, the dark halo being
larger than any directly detectable part of a galaxy, though the two
parts appear to be concentric.

On scales smaller than large-scale structure and larger than galaxies
we then expect to see a transition.  Dark matter should depart 
significantly from luminous matter in its distribution, while retaining
a similar concentration around the densest points.  Exactly how this
happens should tell us something about dark matter itself.

Unfortunately, studying structures in which the density contrast is
highly nonlinear is mathematically difficult.  One approach is to use
galaxy clusters, which should be virialized (or nearly so).  Another
is to find a region in which observational data are abundant and of
high quality, and in which peculiar velocities indicate dynamical
youth.  Here the motions should be simpler than in a dynamically old
system, while at the same time even small motions should be visible.  
Such a region is the
Local Volume, within about 10 megaparsecs (Mpc).  In the Volume there are
radial velocities accurate to one or a few km s$^{-1}$ for hundreds
of galaxies, and (recently) distances good to 10\% or better for
a large fraction.   Using these data, the present study aims to compare
the distribution and motion of luminous matter, in an effort to show
something of the location of gravitating matter.

\section{Data}

\subsection{Sources and Treatment}

Data for 149 galaxies in the Local Volume were collected from the literature and are
presented in Table (\ref{data1}) (ordered by Supergalactic longitude $L$,
as are the rest of the data tables in this paper).  
The main criterion for inclusion was a
distance known to 10\% or better, along with a reliable radial velocity.

The criterion for distance uncertainty was rather restrictive, and in fact it
is possible to derive some kinematic or dynamic properties of the Volume with
a larger set of poorer data (as done in, for instance, \citet{WH03} or 
\citet{KM01}, or any work on large-scale structure).  But the focus here is on the details of the
peculiar velocity field, for which a finer resolution is needed.  A galaxy with
a distance uncertainty of 10\% at 10 Mpc gives a peculiar radial velocity
uncertainty of something over 60 km s$^{-1}$, which is about the largest which
could be tolerated here.  

This means that some popular methods, such as the Tully-Fisher relation, are not
useful, and in fact almost all distances were obtained using Cepheid variables
(Ceph in the table) or the brightness of the tip of the Red Giant Branch as found 
in a $I$, $V-I$ color-magnitude diagram (TRGB).  Some variations on the latter
use different filters ($K$, the SDSS system), or a slightly different calibration
for the absolute magnitude of the tip ($M_I = -4.05$ instead of -4.00); 
they are marked where they occur,
and any resulting differences in distance lie within the quoted errors.  

The technique of surface brightness fluctuation (SBF) seems quite promising and
should have added a few more objects.  However,  the SBF distances
for DDO 181 \citep{HMA03} and ESO 540-32 \citep{JFB98} disagree rather
strongly with the TRGB distances (\citet{KSM02} and \citet{JR01}, respectively),
and NGC 4736 is not much better (\citet{T01} and \citet{KSD03}).  Since the number
of galaxies with only SBF distances is small, it was felt better to leave them out
rather than add a source of unknown error.

        Where more than one distance estimate was available, and
all were consistent, they were combined.  If they were within their
stated relative errors, they were averaged and the uncertainty
reduced by a factor of $\sqrt{N}$; if they were between one and two
uncertainties distant, they were averaged and the larger uncertainty
taken.  If they were not reconcilable (for instance, with IC 10) a
judgement as to the more reliable distance(s) was made.

Even though 149 is a gratifying number of galaxies to work with (being more than four
times the number of good data-points used in a previous calculation, \citet{WH03}),
it still does not include half of the total number present; nor, more importantly,
are all the brightest galaxies there.  To include all galaxies which have been
known or suspected to be brighter than $M_B \sim 18.5$ and in the Local Volume,
21 additional objects are presented in Table (\ref{data2}).  The methods
used in several of these cases might in fact give high-quality data,
since such
things as the brightest-star method and SBF seem to work satisfactorily with larger
galaxies.  This needs to be checked, though, and
Table (\ref{data2}) could usefully be taken as a target list
for a deep TRGB observing program.

Brightest-star distances are assigned an uncertainty of 0.4 magnitude.  Distances to
the major Sculptor Group galaxies were determined by \citet{PC88} from a variety of
methods; as those determined since for NGC 300, 253 and 7793 are systematically
larger, those for NGC 55 and 247 have been adjusted accordingly (note that all of these
are within the uncertainties given in the 1988 paper).  Distances derived solely from
radial velocities use the kinematic isotropic-expansion model derived later in the present paper;
uncertainties are figured as the peculiar radial velocity dispersion times the
effective Hubble constant.  

Apparent $B$ magnitudes are derived from NED\footnote{NED is the NASA/IPAC Extragalactic
Database, which is operated by the Jet Propulsion Laboratory, California Institute of
Technology, under contract with the National Aeronautics and Space Administration.}, 
and are available for almost all objects.
However, most of these numbers come from relatively inaccurate
photographic photometry; for later calculations
they are all assigned an uncertainty of 0.4 magnitude.  Almost all of the $K$ photometry
is from the large, uniform and well-controlled data resulting from the 2MASS survey (here
conveniently extracted also via NED).  It is of course preferable to use better data
where they exist, and for questions of mass-to-light ratios to use near infrared rather
than the much more variable $B$; but less than half of the objects have $K$ measurements.
For most of the following, calculations with be done in both bands, investigating the
effects of sample completeness against photometric accuracy (and assumed closeness to
the actual mass field).  Absolute magnitudes for those dwarf galaxies
without photometry are arbitrarily set at -10 (in $B$ or $K$); these should have no
effect on the results, while avoiding troublesome zeros in calculations.
Galactic absorption is taken from \citet{SFD98} by way of NED except in the case of
the Circinus galaxy, where strong and variable absorption meant that the separate
determination in \citet{FKL77} had to be used.

\subsection{Notes on Particular Objects}

\noindent {\em IC 10}: \\
\indent A distance modulus of $24.9 \pm .2$ is found in \citet{H01}, which is
at variance with the others used here.  Extinction is a large factor
in the disagreement.

\noindent {\em DDO 187}: \\
\indent The Cepheid distance of \citet{HSD98b} is almost three times the
TRGB distance here adopted; this was attributed by \citet{ATK00} to problems with
analyzing a small number of variable stars.

\noindent {\em LMC}: \\
\indent Because the major methods of distance determination (Cepheids
and TRGB) are
calibrated to an LMC distance of 18.50 (50 kpc), it is here fixed at that
distance.

\noindent {\em LMC and SMC}: \\
\indent Neither of these has $K$-band photometry, for reasons which are
obvious on reflection.  They are in fact brighter than most of the
galaxies in the sample and so by rights should be included somehow.
But since they are so close to the Milky Way, their $K$ luminosities do not affect any
calculations separately from it; and their luminosities are almost certainly much smaller than
the {\em uncertainty} in the Milky Way's.  They are therefore ignored in the $K$-band
calculations.

\noindent {\em Phoenix}: \\
\indent There are two mutually incompatible optical radial velocity
measurements for Phoenix, $-52 \pm 6$ \citep{GMG01} and $-13 \pm 9$ 
\citep{IT02a}.  There
is no obvious explanation for the difference between them, nor
any clear reason to choose either.  I have (somewhat arbitrarily) used
the latter.

\noindent {\em PGC 20125 and PGC 39032}: \\
\indent These had no listed radial velocity uncertainty, and the original
measurements could not be traced.  10 km s$^{-1}$ is a conservative educated guess.

\noindent {\em Tucana}: \\
\indent This also had no listed radial velocity uncertainty; the bins in
the paper are 10 km s$^{-1}$ wide, so this is a reasonable amount.

\noindent {\em Carina}: \\
\indent The listed photometry is taken from \citet{MOP93}, and is based on
model-fitting.

\noindent {\em Milky Way}: \\
\indent The apparent magnitudes are back-calculated from the absolute
magnitudes listed in \citet{SSO92}; the uncertainties are guesses.

\noindent {\em NGC 1569}: \\
\indent This is assigned the shorter (and slightly more probable)
TRGB distance from \citet{MK03}.  Due to the uncertainty over which is
the better figure, however, it is not included in
the table of better-quality data.

\noindent {\em NGC 5194 and 5195}: \\
\indent NGC 5195, the smaller part of M 51, has an independent (SBF) distance;
there appears to be no separate determination for NGC 5194, at least of moderate
to high quality.  I have assigned them the same distance.  Normally I strenuously
avoid anything like this ``companion'' or ``member of the same group'' argument
for distance; but in this case there appears to be no choice.  Note that NGC 5194
is not used in any peculiar velocity calculations, only in those of the luminosity
field.

\noindent {\em KKH96, KK98-34, KK98-41, NGC 1560 and KK98-44}: \\
\indent There is no radial velocity error listed for these in the reference.
An estimate of 5 km s$^{-1}$ is here adopted.

\begin{deluxetable}{lrrrrcrrrlrrrrl}
\rotate
\tabletypesize{\scriptsize}
\tablewidth{0pt}
\tablecaption{Local Volume Data \label{data1}}
\tablehead{
\colhead{Name} & \colhead{L} & \colhead{B} & \colhead{RV} & \colhead{$\Delta$RV} & \colhead{RV Source} & 
\colhead{D} & \colhead{$\Delta$D} & \colhead{D Source} & \colhead{D Method} &
\colhead{Absorption} & \colhead{B Mag} & \colhead{K Mag} & \colhead{$\Delta$K Mag} & \colhead{K Mag Source} \\
 & Degrees & Degrees & km s$^{-1}$ & km s$^{-1}$ & & Mpc & Mpc & & &  &  &  &  & }
\startdata
KK98-35 & 10.33 & 0.32 & -66.0 & 5.0 & 1 & 3.16 & 0.32 & 1 & TRGB405 & 2.339 & 17.20 &$\ldots$ & $\ldots$ & $\ldots$ \\ 
IC342 & 10.60 & 0.37 & 31.0 & 3.0 & NED & 3.28 & 0.27 & 2 & Ceph & 2.407 & 9.10 &4.56 & 0.04 & 2M \\ 
UGCA105 & 15.00 & -9.25 & 111.0 & 5.0 & NED & 3.15 & 0.32 & 3 & TRGB & 1.351 & 13.90 &$\ldots$ & $\ldots$ & $\ldots$ \\ 
KK98-44 & 15.04 & -4.21 & 77.0 & 5.0 & 1 & 3.34 & 0.00 & 1 & TRGB405 & 0.936 & 16.10 &$\ldots$ & $\ldots$ & $\ldots$ \\ 
NGC1560 & 16.03 & 0.79 & -36.0 & 5.0 & 1 & 3.45 & 0.37 & 1 & TRGB405 & 0.812 & 12.16 &8.93 & 0.03 & 2M \\ 
KK98-41 & 16.09 & 1.87 & -47.0 & 5.0 & 1 & 3.93 & 0.47 & 1 & TRGB405 & 0.951 & 14.84 &$\ldots$ & $\ldots$ & $\ldots$ \\ 
KKH34 & 22.54 & -0.41 & 110.0 & 2.0 & NED & 4.61 & 0.36 & 4 & TRGB405 & 1.076 & 17.10 &$\ldots$ & $\ldots$ & $\ldots$ \\ 
NGC6789 & 23.28 & 41.59 & -141.0 & 9.0 & NED & 3.63 & 0.37 & 5 & TRGB & 0.302 & 13.76 &12.24 & 0.09 & 2M \\ 
NGC2366 & 29.46 & -4.86 & 100.0 & 3.0 & NED & 3.19 & 0.41 & 6 & TRGB405 & 0.157 & 11.43 &10.62 & 0.13 & 2M \\ 
NGC2403 & 30.80 & -8.31 & 131.0 & 3.0 & NED & 3.16 & 0.25 & 7,8 & TRGBsdss,Ceph & 0.172 & 8.93 &6.19 & 0.04 & 2M \\ 
NGC6503 & 33.14 & 34.63 & 60.0 & 7.0 & NED & 5.27 & 0.77 & 4 & TRGB405 & 0.138 & 10.91 &7.30 & 0.02 & 2M \\ 
DDO50 & 33.26 & -2.36 & 157.0 & 1.0 & NED & 3.31 & 0.15 & 6,9 & TRGB405,Ceph & 0.139 & 11.10 &8.79 & 0.04 & 2M \\ 
KDG52 & 33.52 & -1.93 & 113.0 & 5.0 & NED & 3.55 & 0.26 & 6 & TRGB405 & 0.091 & 16.50 &$\ldots$ & $\ldots$ & $\ldots$ \\ 
UGC4483 & 35.02 & -2.65 & 156.0 & 5.0 & NED & 3.29 & 0.13 & 10,11 & TRGB & 0.140 & 15.12 &$\ldots$ & $\ldots$ & $\ldots$ \\ 
DDO53 & 36.25 & -6.04 & 19.0 & 10.0 & NED & 3.56 & 0.25 & 6 & TRGB405 & 0.160 & 14.48 &$\ldots$ & $\ldots$ & $\ldots$ \\ 
UGC3755 & 36.82 & -63.36 & 314.0 & 4.0 & NED & 5.10 & 0.27 & 12,4 & TRGB & 0.384 & 14.10 &$\ldots$ & $\ldots$ & $\ldots$ \\ 
VIIZw403 & 36.87 & 11.40 & -100.0 & 1.0 & NED & 4.41 & 0.10 & 12,13 & TRGB & 0.155 & 14.50 &12.60 & 0.10 & 14 \\ 
HoI & 38.77 & 1.33 & 143.0 & 1.0 & NED & 3.84 & 0.46 & 6 & TRGB405 & 0.207 & 13.00 &$\ldots$ & $\ldots$ & $\ldots$ \\ 
M82 & 40.72 & 1.05 & 203.0 & 4.0 & NED & 3.89 & 0.38 & 15 & TRGB & 0.685 & 9.30 &4.66 & 0.01 & 2M \\ 
BK3N & 41.07 & 0.41 & -40.0 & 0.0 & NED & 4.29 & 0.26 & 6 & TRGB405 & 0.385 & 17.10 &$\ldots$ & $\ldots$ & $\ldots$ \\ 
M81 & 41.12 & 0.59 & -34.0 & 4.0 & NED & 3.55 & 0.13 & 8 & Ceph & 0.346 & 7.89 &3.83 & 0.02 & 2M \\ 
NGC2976 & 41.33 & -0.78 & 3.0 & 5.0 & NED & 3.56 & 0.38 & 6 & TRGB405 & 0.300 & 10.82 &7.52 & 0.03 & 2M \\ 
KK98-81 & 41.54 & 0.33 & -135.0 & 30.0 & NED & 3.60 & 0.25 & 16 & TRGB & 0.309 & 15.60 &$\ldots$ & $\ldots$ & $\ldots$ \\ 
NGC3077 & 41.85 & 0.83 & 14.0 & 4.0 & NED & 3.85 & 0.37 & 17 & TRGB & 0.289 & 10.61 &7.30 & 0.02 & 2M \\ 
Garland & 41.91 & 0.83 & 50.0 & 0.0 & NED & 3.82 & 0.42 & 6 & TRGB405 & 0.289 & 0.00 &$\ldots$ & $\ldots$ & $\ldots$ \\ 
DDO82 & 42.04 & 3.85 & 180.0 & 60.0 & NED & 4.00 & 0.41 & 6 & TRGB405 & 0.188 & 13.47 &$\ldots$ & $\ldots$ & $\ldots$ \\ 
KK98-85 & 42.74 & 0.44 & -18.0 & 14.0 & NED & 3.70 & 0.26 & 16 & TRGB & 0.235 & 18.00 &$\ldots$ & $\ldots$ & $\ldots$ \\ 
DDO71 & 43.51 & -0.58 & -126.0 & 5.0 & NED & 3.50 & 0.24 & 16 & TRGB & 0.412 & 18.00 &$\ldots$ & $\ldots$ & $\ldots$ \\ 
IC2574 & 43.63 & 2.31 & 57.0 & 2.0 & NED & 4.02 & 0.42 & 6 & TRGB405 & 0.156 & 10.80 &10.72 & 0.10 & 2M \\ 
Draco & 43.79 & 44.23 & -292.0 & 21.0 & NED & 0.76 & 0.01 & 18 & RRL & 0.118 & 10.90 &$\ldots$ & $\ldots$ & $\ldots$ \\ 
KDG73 & 44.03 & 4.75 & -132.0 & 6.0 & NED & 3.82 & 0.23 & 6 & TRGB405 & 0.080 & 14.90 &$\ldots$ & $\ldots$ & $\ldots$ \\ 
DDO78 & 44.10 & 1.69 & 55.0 & 10.0 & 19 & 3.72 & 0.26 & 16 & TRGB & 0.094 & 15.80 &$\ldots$ & $\ldots$ & $\ldots$ \\ 
DDO47 & 46.54 & -55.49 & 272.0 & 1.0 & NED & 5.18 & 0.60 & 4 & TRGB405 & 0.145 & 13.60 &$\ldots$ & $\ldots$ & $\ldots$ \\ 
KK98-65 & 46.93 & -55.67 & 407.0 & 10.0 & 20 & 4.51 & 0.37 & 4 & TRGB405 & 0.137 & 15.60 &$\ldots$ & $\ldots$ & $\ldots$ \\ 
NGC4236 & 47.11 & 11.38 & 0.0 & 4.0 & NED & 4.45 & 0.45 & 6 & TRGB405 & 0.063 & 10.05 &9.01 & 0.05 & 2M \\ 
UMi & 47.71 & 27.10 & -247.0 & 1.0 & NED & 0.08 & 0.00 & 21 & CMD & 0.137 & 11.90 &$\ldots$ & $\ldots$ & $\ldots$ \\ 
DDO165 & 49.61 & 15.58 & 37.0 & 8.0 & NED & 4.57 & 0.40 & 6 & TRGB405 & 0.104 & 12.80 &$\ldots$ & $\ldots$ & $\ldots$ \\ 
UGC4115 & 54.21 & -56.22 & 338.0 & 5.0 & NED & 5.50 & 0.58 & 4 & TRGB405 & 0.122 & 15.20 &$\ldots$ & $\ldots$ & $\ldots$ \\ 
KKR25 & 56.09 & 40.37 & -135.0 & 2.0 & 22 & 1.86 & 0.12 & 23 & TRGB403 & 0.036 & 17.00 &$\ldots$ & $\ldots$ & $\ldots$ \\ 
NGC5204 & 59.40 & 17.85 & 201.0 & 1.0 & NED & 4.65 & 0.53 & 24 & TRGB & 0.054 & 11.73 &9.51 & 0.06 & 2M \\ 
NGC3738 & 59.57 & 1.79 & 229.0 & 4.0 & NED & 4.90 & 0.54 & 24 & TRGB & 0.045 & 12.13 &$\ldots$ & $\ldots$ & $\ldots$ \\ 
UGC8508 & 63.09 & 17.91 & 62.0 & 5.0 & NED & 2.56 & 0.15 & 3 & TRGB & 0.064 & 14.40 &$\ldots$ & $\ldots$ & $\ldots$ \\ 
M101 & 63.58 & 22.61 & 241.0 & 2.0 & NED & 6.83 & 0.31 & 25,8 & Ceph & 0.037 & 8.31 &5.51 & 0.05 & 2M \\ 
UGC7298 & 63.89 & 6.63 & 172.0 & 5.0 & NED & 4.21 & 0.32 & 24 & TRGB & 0.098 & 15.00 &$\ldots$ & $\ldots$ & $\ldots$ \\ 
UGC6541 & 64.19 & -0.79 & 249.0 & 4.0 & NED & 3.89 & 0.47 & 24 & TRGB & 0.078 & 0.00 &$\ldots$ & $\ldots$ & $\ldots$ \\ 
NGC3741 & 67.96 & -2.08 & 229.0 & 4.0 & NED & 3.03 & 0.33 & 24 & TRGB & 0.110 & 14.30 &$\ldots$ & $\ldots$ & $\ldots$ \\ 
M106 & 68.74 & 5.55 & 448.0 & 3.0 & NED & 7.47 & 0.28 & 8,26 & Ceph,Geo & 0.069 & 9.10 &5.46 & 0.02 & 2M \\ 
LeoA & 69.91 & -25.80 & 20.0 & 4.0 & NED & 0.74 & 0.06 & 27,28 & CMD,RRL & 0.089 & 12.92 &$\ldots$ & $\ldots$ & $\ldots$ \\ 
KK98-109 & 70.24 & -0.92 & 225.0 & 10.0 & 20 & 4.51 & 0.34 & 24 & TRGB & 0.083 & 17.50 &$\ldots$ & $\ldots$ & $\ldots$ \\ 
DDO167 & 71.75 & 14.42 & 164.0 & 6.0 & NED & 4.19 & 0.47 & 24 & TRGB & 0.041 & 17.00 &$\ldots$ & $\ldots$ & $\ldots$ \\ 
DDO168 & 72.19 & 14.55 & 192.0 & 1.0 & NED & 4.33 & 0.49 & 24 & TRGB & 0.065 & 12.70 &$\ldots$ & $\ldots$ & $\ldots$ \\ 
NGC4449 & 72.30 & 6.18 & 207.0 & 4.0 & NED & 4.21 & 0.50 & 24 & TRGB & 0.083 & 9.99 &7.25 & 0.04 & 2M \\ 
DDO125 & 72.85 & 5.93 & 196.0 & 4.0 & NED & 2.51 & 0.11 & 3,12 & TRGB & 0.088 & 12.84 &$\ldots$ & $\ldots$ & $\ldots$ \\ 
DDO190 & 74.07 & 26.85 & 150.0 & 4.0 & NED & 2.85 & 0.16 & 3,29 & TRGB & 0.052 & 13.25 &$\ldots$ & $\ldots$ & $\ldots$ \\ 
DDO99 & 74.93 & -2.12 & 242.0 & 1.0 & NED & 2.64 & 0.21 & 3 & TRGB & 0.113 & 13.40 &$\ldots$ & $\ldots$ & $\ldots$ \\ 
M94 & 76.24 & 9.50 & 308.0 & 1.0 & NED & 4.66 & 0.59 & 24 & TRGB & 0.076 & 8.99 &5.11 & 0.02 & 2M \\ 
NGC4244 & 77.73 & 2.41 & 243.0 & 4.0 & NED & 4.49 & 0.47 & 24 & TRGB & 0.090 & 10.88 &7.72 & 0.05 & 2M \\ 
UGCA290 & 77.94 & 6.41 & 445.0 & 9.0 & NED & 6.61 & 0.37 & 30 & TRGB & 0.060 & 0.00 &$\ldots$ & $\ldots$ & $\ldots$ \\ 
DDO181 & 78.09 & 18.58 & 201.0 & 1.0 & NED & 3.01 & 0.29 & 3 & TRGB & 0.026 & 14.70 &$\ldots$ & $\ldots$ & $\ldots$ \\ 
IC3687 & 78.42 & 7.27 & 354.0 & 1.0 & NED & 4.57 & 0.48 & 24 & TRGB & 0.088 & 13.71 &$\ldots$ & $\ldots$ & $\ldots$ \\ 
DDO126 & 78.93 & 4.03 & 218.0 & 5.0 & NED & 4.87 & 0.55 & 24 & TRGB & 0.060 & 14.20 &$\ldots$ & $\ldots$ & $\ldots$ \\ 
NGC4214 & 79.02 & 1.60 & 291.0 & 3.0 & NED & 2.82 & 0.17 & 31,32 & TRGB & 0.094 & 10.24 &7.91 & 0.05 & 2M \\ 
DDO113 & 79.08 & 1.43 & 284.0 & 6.0 & NED & 2.86 & 0.14 & 3 & TRGB & 0.090 & 15.40 &$\ldots$ & $\ldots$ & $\ldots$ \\ 
IC4182 & 80.34 & 11.61 & 321.0 & 1.0 & NED & 4.53 & 0.13 & 8 & Ceph & 0.059 & 13.00 &$\ldots$ & $\ldots$ & $\ldots$ \\ 
UGC7605 & 80.39 & 3.92 & 309.0 & 5.0 & NED & 4.43 & 0.53 & 24 & TRGB & 0.063 & 15.00 &$\ldots$ & $\ldots$ & $\ldots$ \\ 
NGC4395 & 82.31 & 2.73 & 319.0 & 1.0 & NED & 4.46 & 0.34 & 24,33 & TRGB,Ceph & 0.074 & 10.64 &9.98 & 0.06 & 2M \\ 
UGC8833 & 83.54 & 21.09 & 228.0 & 5.0 & NED & 3.19 & 0.21 & 24 & TRGB & 0.051 & 16.50 &$\ldots$ & $\ldots$ & $\ldots$ \\ 
LeoI & 88.90 & -34.56 & 285.0 & 2.0 & NED & 0.26 & 0.01 & 12 & TRGB & 0.156 & 11.18 &$\ldots$ & $\ldots$ & $\ldots$ \\ 
M95 & 94.09 & -27.08 & 778.0 & 4.0 & NED & 9.33 & 0.39 & 8 & Ceph & 0.120 & 10.53 &6.67 & 0.04 & 2M \\ 
M96 & 94.29 & -26.41 & 897.0 & 4.0 & NED & 9.86 & 0.27 & 8 & Ceph & 0.109 & 10.11 &6.32 & 0.02 & 2M \\ 
SextansB & 95.46 & -39.62 & 301.0 & 1.0 & NED & 1.33 & 0.04 & 12,3,34 & TRGB & 0.137 & 11.85 &$\ldots$ & $\ldots$ & $\ldots$ \\ 
M66 & 96.58 & -18.42 & 727.0 & 3.0 & NED & 9.38 & 0.37 & 8 & Ceph & 0.140 & 9.65 &5.88 & 0.02 & 2M \\ 
DDO187 & 97.83 & 24.35 & 154.0 & 4.0 & NED & 2.50 & 0.20 & 35 & TRGB & 0.105 & 14.38 &$\ldots$ & $\ldots$ & $\ldots$ \\ 
GR8 & 102.98 & 4.67 & 214.0 & 3.0 & NED & 2.22 & 0.26 & 36,37 & TRGB & 0.113 & 14.68 &$\ldots$ & $\ldots$ & $\ldots$ \\ 
Sextans & 105.39 & -39.23 & 224.0 & 2.0 & NED & 0.10 & 0.00 & 38 & CMD & 0.215 & 12.00 &$\ldots$ & $\ldots$ & $\ldots$ \\ 
SextansA & 109.01 & -40.66 & 324.0 & 1.0 & NED & 1.32 & 0.04 & 39 & Ceph,TRGB400 & 0.188 & 11.86 &$\ldots$ & $\ldots$ & $\ldots$ \\ 
KKH86 & 116.34 & 15.47 & 283.0 & 3.0 & NED & 2.61 & 0.16 & 3 & TRGB & 0.118 & 16.80 &$\ldots$ & $\ldots$ & $\ldots$ \\ 
NGC3109 & 137.96 & -45.10 & 403.0 & 1.0 & NED & 1.32 & 0.03 & 12,3,40,41 & TRGB,TRGB,Ceph,TRGB & 0.288 & 10.39 &9.28 & 0.07 & 2M \\ 
Antlia & 139.93 & -44.80 & 351.0 & 15.0 & 42 & 1.32 & 0.10 & 43 & TRGB & 0.342 & 16.19 &$\ldots$ & $\ldots$ & $\ldots$ \\ 
NGC3621 & 145.65 & -28.57 & 727.0 & 5.0 & NED & 6.43 & 0.31 & 44,8 & Ceph & 0.346 & 10.18 &6.60 & 0.04 & 2M \\ 
PGC48111 & 146.23 & 1.64 & 587.0 & 4.0 & NED & 4.61 & 0.40 & 45 & TRGB & 0.295 & 14.96 &$\ldots$ & $\ldots$ & $\ldots$ \\ 
KK98-112 & 146.93 & -20.99 & 640.0 & 5.0 & NED & 5.22 & 0.43 & 45 & TRGB & 0.323 & 16.60 &$\ldots$ & $\ldots$ & $\ldots$ \\ 
IC4316 & 147.25 & 1.98 & 674.0 & 53.0 & NED & 4.41 & 0.47 & 45 & TRGB & 0.237 & 14.97 &$\ldots$ & $\ldots$ & $\ldots$ \\ 
KK98-208 & 147.62 & 1.00 & 381.0 & 0.0 & NED & 4.68 & 0.41 & 45 & TRGB & 0.192 & 14.30 &$\ldots$ & $\ldots$ & $\ldots$ \\ 
M83 & 147.93 & 0.99 & 516.0 & 4.0 & NED & 4.27 & 0.31 & 46 & Ceph & 0.284 & 8.20 &4.62 & 0.03 & 2M \\ 
KK98-200 & 148.16 & -1.90 & 485.0 & 3.0 & NED & 4.63 & 0.31 & 45 & TRGB & 0.298 & 16.67 &$\ldots$ & $\ldots$ & $\ldots$ \\ 
NGC5264 & 148.30 & 1.92 & 478.0 & 3.0 & NED & 4.53 & 0.55 & 45 & TRGB & 0.223 & 12.60 &10.54 & 0.05 & 2M \\ 
NGC5253 & 149.81 & 1.01 & 404.0 & 4.0 & NED & 3.25 & 0.21 & 8 & Ceph & 0.242 & 10.87 &8.25 & 0.04 & 2M \\ 
PGC39032 & 152.26 & -17.62 & 613.0 & 10.0 & NED & 3.18 & 0.25 & 45 & TRGB & 0.400 & 15.16 &$\ldots$ & $\ldots$ & $\ldots$ \\ 
NGC5102 & 153.42 & -4.07 & 467.0 & 7.0 & NED & 3.40 & 0.39 & 45 & TRGB & 0.237 & 10.35 &6.92 & 0.04 & 2M \\ 
PGC47171 & 158.39 & -4.42 & 513.0 & 6.0 & NED & 3.73 & 0.43 & 45 & TRGB & 0.466 & 12.90 &$\ldots$ & $\ldots$ & $\ldots$ \\ 
NGC5128 & 159.75 & -5.25 & 547.0 & 5.0 & NED & 3.75 & 0.28 & 47 & TRGBK & 0.496 & 7.84 &3.94 & 0.02 & 2M \\ 
PGC48738 & 159.79 & -1.46 & 541.0 & 4.0 & NED & 3.40 & 0.39 & 45 & TRGB & 0.376 & 13.99 &$\ldots$ & $\ldots$ & $\ldots$ \\ 
NGC5408 & 160.59 & 1.90 & 509.0 & 7.0 & NED & 4.81 & 0.38 & 45 & TRGB & 0.298 & 12.20 &11.39 & 0.14 & 2M \\ 
PGC51659 & 166.95 & 3.84 & 397.0 & 68.0 & NED & 3.58 & 0.33 & 45 & TRGB & 0.561 & 16.50 &$\ldots$ & $\ldots$ & $\ldots$ \\ 
MilkyWay & 185.79 & 42.31 & -9.0 & 5.0 & 48 & 0.01 & 0.00 & 49 & various & 0.000 & -5.98 &-9.18 & 0.50 & 50,51 \\ 
IC3104 & 195.83 & -17.06 & 430.0 & 5.0 & NED & 2.27 & 0.19 & 3 & TRGB & 1.701 & 13.63 &13.61 & 0.35 & 2M \\ 
NGC2915 & 197.37 & -26.06 & 468.0 & 5.0 & NED & 3.78 & 0.45 & 4 & TRGB405 & 1.185 & 13.25 &9.83 & 0.06 & 2M \\ 
PGC20125 & 204.47 & -47.15 & 554.0 & 10.0 & NED & 4.90 & 0.50 & 4 & TRGB405 & 0.513 & 14.20 &$\ldots$ & $\ldots$ & $\ldots$ \\ 
Carina & 210.10 & -54.64 & 229.0 & 60.0 & NED & 0.09 & 0.01 & 52 & CMD & 0.271 & 10.80 &$\ldots$ & $\ldots$ & $\ldots$ \\ 
LMC & 215.80 & -34.12 & 278.0 & 2.0 & NED & 0.05 & 0.00 & fixed & various & 0.324 & 0.90 &$\ldots$ & $\ldots$ & $\ldots$ \\ 
SagDIG & 221.27 & 55.52 & -77.0 & 4.0 & NED & 1.05 & 0.05 & 3,53 & TRGB & 0.522 & 15.50 &$\ldots$ & $\ldots$ & $\ldots$ \\ 
SMC & 224.23 & -14.83 & 158.0 & 4.0 & NED & 0.06 & 0.00 & 54 & EB & 0.160 & 2.70 &$\ldots$ & $\ldots$ & $\ldots$ \\ 
PGC19337 & 226.90 & -78.97 & 499.0 & 1.0 & NED & 4.23 & 0.45 & 4 & TRGB405 & 0.337 & 13.69 &$\ldots$ & $\ldots$ & $\ldots$ \\ 
Tucana & 227.61 & -0.92 & 182.0 & 10.0 & 55 & 0.87 & 0.06 & 56 & TRGB & 0.137 & 15.70 &$\ldots$ & $\ldots$ & $\ldots$ \\ 
NGC1313 & 227.98 & -28.22 & 475.0 & 3.0 & NED & 4.13 & 0.11 & 12 & TRGB & 0.471 & 9.20 &7.57 & 0.06 & 2M \\ 
NGC6822 & 229.08 & 57.10 & -57.0 & 2.0 & NED & 0.48 & 0.02 & 57,58 & TRGB,RRL & 1.020 & 9.31 &6.72 & 0.05 & 2M \\ 
NGC1705 & 231.83 & -45.54 & 628.0 & 9.0 & NED & 5.11 & 0.61 & 59 & TRGB & 0.035 & 12.77 &10.53 & 0.06 & 2M \\ 
IC5152 & 234.23 & 11.53 & 124.0 & 3.0 & NED & 1.88 & 0.20 & 60,3 & TRGB & 0.106 & 11.06 &9.28 & 0.03 & 2M \\ 
PGC09962 & 235.18 & -25.96 & 513.0 & 9.0 & NED & 4.66 & 0.47 & 4 & TRGB405 & 0.110 & 13.24 &13.57 & 0.19 & 2M \\ 
KK98-54 & 237.66 & -77.50 & 495.0 & 22.0 & NED & 4.99 & 0.60 & 4 & TRGB405 & 0.283 & 15.60 &$\ldots$ & $\ldots$ & $\ldots$ \\ 
DDO210 & 252.08 & 50.24 & -140.7 & 1.3 & 61 & 0.94 & 0.03 & 62,3 & TRGB & 0.221 & 14.00 &$\ldots$ & $\ldots$ & $\ldots$ \\ 
Phoenix & 254.29 & -20.86 & -13.0 & 9.0 & 88 & 0.40 & 0.01 & 63,64 & TRGB & 0.067 & 13.07 &$\ldots$ & $\ldots$ & $\ldots$ \\ 
PGC01641 & 254.37 & -5.27 & 117.0 & 5.0 & NED & 1.92 & 0.09 & 3 & TRGB & 0.024 & 15.60 &$\ldots$ & $\ldots$ & $\ldots$ \\ 
ESO245-05 & 255.14 & -19.74 & 395.0 & 6.0 & NED & 4.43 & 0.45 & 65 & TRGB405 & 0.069 & 12.70 &$\ldots$ & $\ldots$ & $\ldots$ \\ 
NGC625 & 257.27 & -17.74 & 396.0 & 1.0 & NED & 3.89 & 0.22 & 66 & TRGB & 0.071 & 11.71 &9.09 & 0.04 & 2M \\ 
UGCA438 & 258.88 & 9.28 & 62.0 & 5.0 & NED & 2.16 & 0.10 & 3,67 & TRGB & 0.064 & 13.86 &$\ldots$ & $\ldots$ & $\ldots$ \\ 
NGC300 & 259.81 & -9.50 & 144.0 & 1.0 & NED & 2.02 & 0.07 & 8 & Ceph & 0.055 & 8.95 &6.38 & 0.06 & 2M \\ 
UGCA442 & 260.78 & 6.12 & 267.0 & 8.0 & NED & 4.27 & 0.52 & 65 & TRGB405 & 0.072 & 13.60 &12.77 & 0.33 & 2M \\ 
NGC7793 & 261.30 & 3.12 & 230.0 & 2.0 & NED & 3.91 & 0.41 & 65 & TRGB405 & 0.084 & 9.98 &6.86 & 0.06 & 2M \\ 
Sculptor & 263.98 & -9.68 & 110.0 & 1.0 & NED & 0.08 & 0.00 & 68 & RRL & 0.077 & 10.05 &$\ldots$ & $\ldots$ & $\ldots$ \\ 
Fornax & 265.37 & -30.27 & 53.0 & 9.0 & NED & 0.14 & 0.01 & 69 & RRL & 0.087 & 9.28 &$\ldots$ & $\ldots$ & $\ldots$ \\ 
NGC253 & 271.57 & -5.01 & 241.0 & 2.0 & NED & 3.94 & 0.37 & 65 & TRGB405 & 0.081 & 8.04 &3.77 & 0.02 & 2M \\ 
IC1574 & 274.23 & -3.21 & 361.0 & 7.0 & NED & 4.92 & 0.58 & 65 & TRGB405 & 0.065 & 15.09 &$\ldots$ & $\ldots$ & $\ldots$ \\ 
DDO6 & 275.84 & -4.40 & 301.0 & 5.0 & NED & 3.34 & 0.24 & 65 & TRGB405 & 0.073 & 15.19 &$\ldots$ & $\ldots$ & $\ldots$ \\ 
WLM & 277.81 & 8.09 & -116.0 & 2.0 & NED & 0.89 & 0.04 & 70 & TRGB & 0.160 & 11.03 &$\ldots$ & $\ldots$ & $\ldots$ \\ 
IC1613 & 299.15 & -1.78 & -234.0 & 1.0 & NED & 0.73 & 0.02 & 8,71,72 & Ceph,various,TRGB & 0.108 & 9.88 &12.77 & 0.11 & 2M \\ 
DDO216 & 305.83 & 24.30 & -183.3 & 1.3 & 61 & 0.76 & 0.10 & 73 & CMD & 0.284 & 13.21 &$\ldots$ & $\ldots$ & $\ldots$ \\ 
UGC685 & 313.34 & 1.61 & 157.0 & 1.0 & NED & 4.79 & 0.30 & 31 & TRGB & 0.246 & 14.23 &$\ldots$ & $\ldots$ & $\ldots$ \\ 
AndVI & 317.31 & 20.56 & -354.0 & 3.0 & NED & 0.77 & 0.03 & 74,75 & TRGB,Cep & 0.276 & 14.17 &$\ldots$ & $\ldots$ & $\ldots$ \\ 
LGS3 & 318.13 & 3.80 & -277.0 & 5.0 & NED & 0.62 & 0.02 & 76 & TRGB,HB,RC & 0.177 & 14.20 &$\ldots$ & $\ldots$ & $\ldots$ \\ 
KK98-16 & 327.50 & -5.41 & 205.0 & 10.0 & 20 & 4.74 & 0.53 & 4 & TRGB405 & 0.301 & 16.30 &$\ldots$ & $\ldots$ & $\ldots$ \\ 
M33 & 328.47 & -0.09 & -179.0 & 3.0 & NED & 0.80 & 0.02 & 8,77,78 & Ceph,TRGBClump,Ceph & 0.181 & 6.27 &4.10 & 0.04 & 2M \\ 
KK98-17 & 328.71 & -6.08 & 155.0 & 10.0 & 20 & 4.72 & 0.41 & 4 & TRGB405 & 0.239 & 17.20 &$\ldots$ & $\ldots$ & $\ldots$ \\ 
AndII & 330.01 & 4.23 & -188.0 & 3.0 & NED & 0.66 & 0.01 & 79,80,78 & RRL,TRGB & 0.269 & 13.50 &$\ldots$ & $\ldots$ & $\ldots$ \\ 
AndIII & 331.13 & 13.10 & -351.0 & 9.0 & NED & 0.76 & 0.07 & 81 & TRGB & 0.244 & 15.00 &$\ldots$ & $\ldots$ & $\ldots$ \\ 
NGC404 & 331.85 & 6.25 & -48.0 & 9.0 & NED & 3.20 & 0.21 & 82,3 & TRGB & 0.253 & 11.21 &8.57 & 0.03 & 89 \\ 
KKH98 & 332.35 & 23.17 & -136.0 & 3.0 & NED & 2.45 & 0.12 & 3 & TRGB & 0.532 & 16.70 &$\ldots$ & $\ldots$ & $\ldots$ \\ 
AndI & 333.05 & 11.41 & -368.0 & 11.0 & NED & 0.76 & 0.03 & 80,78 & TRGB & 0.234 & 13.60 &$\ldots$ & $\ldots$ & $\ldots$ \\ 
NGC925 & 335.45 & -9.47 & 553.0 & 3.0 & NED & 9.12 & 0.17 & 8 & Ceph & 0.326 & 10.69 &8.74 & 0.03 & 2M \\ 
M31 & 336.19 & 12.55 & -300.0 & 4.0 & NED & 0.75 & 0.02 & 8 & Ceph & 0.268 & 4.36 &0.98 & 0.02 & 2M \\ 
NGC205 & 336.54 & 13.06 & -241.0 & 3.0 & NED & 0.72 & 0.07 & 83 & TRGB & 0.268 & 8.92 &5.59 & 0.05 & 2M \\ 
KKH18 & 339.26 & -15.93 & 216.0 & 3.0 & NED & 4.43 & 0.49 & 4 & TRGB405 & 0.860 & 16.70 &$\ldots$ & $\ldots$ & $\ldots$ \\ 
NGC185 & 343.27 & 14.30 & -202.0 & 3.0 & NED & 0.62 & 0.06 & 84 & TRGB & 0.787 & 10.10 &6.56 & 0.05 & 2M \\ 
NGC147 & 343.32 & 15.27 & -193.0 & 3.0 & NED & 0.76 & 0.10 & 85 & RGB/HB & 0.747 & 10.47 &7.20 & 0.06 & 2M \\ 
KKH96 & 345.65 & 26.05 & -307.0 & 5.0 & 1 & 0.79 & 0.04 & 1 & TRGB405 & 0.846 & 13.60 &$\ldots$ & $\ldots$ & $\ldots$ \\ 
KKH5 & 347.21 & 10.31 & 61.0 & 2.0 & NED & 4.27 & 0.33 & 4 & TRGB405 & 1.218 & 17.10 &$\ldots$ & $\ldots$ & $\ldots$ \\ 
IC10 & 354.43 & 17.88 & -348.0 & 1.0 & NED & 0.66 & 0.08 & 86 & Ceph,TRGB & 6.588 & 11.80 &6.01 & 0.03 & 2M \\ 
MaffeiI & 359.29 & 1.44 & 66.4 & 5.0 & 87 & 3.01 & 0.30 & 87 & various & 6.120 & 13.47 &4.68 & 0.02 & 2M \\ 
\enddata
\tablerefs{(1) \citet{KSDG3}; (2) \citet{SCH02}; (3) \citet{KSM02}; (4) \citet{KMS03};
(5) \citet{DSH01}; (6) \citet{KDG02}; (7) \citet{D03}; (8) \citet{FMG01}; 
(9) \citet{HSD98}; (10) \citet{IT02b}; (11) \citet{Do01}; (12) \citet{MDM02}; 
(13) \citet{LTO98}; (14) \citet{TBD81}; (15) \citet{SM99}; (16) \citet{K00}; (17) \citet{SM01};
(18) \citet{BSS03}; (19) \citet{SKB01}; (20) \citet{HKK00}; (21) \citet{BFO02}; 
(22) \citet{HKK00b}; (23) \citet{K01a}; (24) \citet{KSD03}; (25) \citet{S98};
(26) \citet{HMG99}; (27) \citet{TGC98}; (28) \citet{D02}; (29) \citet{AT00}; (30) \citet{CSH00};
(31) \citet{MCM02}; (32) \citet{DSH02}; (33) \citet{THS04}; (34) \citet{SMF97};
(35) \citet{ATK00}; (36) \citet{DSG98}; (37) \citet{T95}; (38) \citet{L03}; (39) \citet{Do03};
(40) \citet{MPC97}; (41) \citet{HSD98b}; (42) \citet{TI00}; (43) \citet{A97};
(44) \citet{R97}; (45) \citet{KSD02}; (46) \citet{TTS03}; (47) \citet{R03}; (48) \citet{MB};
(49) \citet{LB}; (50) \citet{SSO92}; (51) \citet{KDF91}; (52) \citet{M97}; (53) \citet{KAM99};
(54) \citet{HHH03}; (55) \citet{TIC04}; (56) \citet{SHP96}; (57) \citet{W03}; 
(58) \citet{CHB03}; (59) \citet{TSB01}; (60) \citet{ZM99}; (61) \citet{YVL03};
(62) \citet{LA99}; (63) \citet{MGA99}; (64) \citet{HSM99}; (65) \citet{KGS03};
(66) \citet{CDS03}; (67) \citet{LB99}; (68) \citet{KD77}; (69) \citet{MG03}; (70) \citet{MZ97};
(71) \citet{D01}; (72) \citet{CTG99}; (73) \citet{GTD98}; (74) \citet{HSG99};
(75) \citet{PAJ02}; (76) \citet{MDL01}; (77) \citet{LKS02}; (78) \citet{MIF04};
(79) \citet{PAJ03}; (80) \citet{DAC00}; (81) \citet{ADC93}; (82) \citet{TGA03};
(83) \citet{MKD84}; (84) \citet{LFM93}; (85) \citet{HHG97}; (86) \citet{SMF99};
(87) \citet{F03}; (88) \citet{IT02a}; (89) \citet{FRA78}}
\tablecomments{Observational data on Local Volume galaxies, derived from the literature.  NED is the
NASA Extragalactic Database (http://nedwww.ipac.caltech.edu/index.html); 2M, the 2MASS
survey, data extracted through NED.  TRGB, tip of the Red Giant Branch (sometimes using the SDSS
filters or $K$, and noted as such); HB, horizonal branch; CMD, color-magnitude diagram; RRL, RR Lyraes;
Ceph, Cepheids; Clump or RC, red clump; EB, eclipsing binaries; Geo, a Geometric method using
masers.  A lack of available data is denoted by ``$\ldots$''.}
\end{deluxetable}

\begin{deluxetable}{lrrrrcrrrlrrrrl}
\rotate
\tabletypesize{\scriptsize}
\tablecaption{Additional Local Volume Data \label{data2}}
\tablehead{
\colhead{Name} & \colhead{L} & \colhead{B} & \colhead{RV} & \colhead{$\Delta$RV} & \colhead{RV Source} &
\colhead{D} & \colhead{$\Delta$D} & \colhead{D Source} & \colhead{D Method} &
\colhead{Absorption} & \colhead{B Mag} & \colhead{K Mag} & \colhead{$\Delta$K Mag} & \colhead{K Mag Source} \\
& Degrees & Degrees & km s$^{-1}$ & km s$^{-1}$ & & Mpc & Mpc & & &  &  &  &  & }
\startdata
NGC6946 & 10.03 & 42.00 & 48.0 & 2.0 & NED & 6.80 & 1.40 & 1 & stars & 1.475 & 9.61 &5.37 & 0.03 & 2M \\ 
NGC1569 & 11.91 & -4.92 & -104.0 & 4.0 & NED & 1.95 & 0.20 & 2 & TRGB* & 3.020 & 11.86 &7.86 & 0.02 & 2M \\ 
NGC2683 & 55.87 & -33.42 & 411.0 & 1.0 & NED & 9.20 & 1.80 & 3 & stars & 0.142 & 10.60 &6.33 & 0.02 & 2M \\ 
NGC5585 & 60.40 & 24.67 & 305.0 & 3.0 & NED & 8.70 & 1.70 & 3 & stars & 0.067 & 11.20 &9.50 & 0.05 & 2M \\ 
NGC5195 & 71.09 & 17.35 & 465.0 & 10.0 & NED & 7.66 & 0.95 & 4 & SBF & 0.155 & 10.45 &6.25 & 0.03 & 2M \\ 
NGC5194 & 71.17 & 17.32 & 463.0 & 3.0 & NED & 7.66 & 0.95 & 4 & SBF & 0.150 & 8.96 &5.50 & 0.03 & 2M \\ 
NGC2903 & 73.53 & -36.44 & 556.0 & 1.0 & NED & 8.90 & 1.80 & 3 & stars & 0.134 & 9.68 &6.04 & 0.02 & 2M \\ 
NGC3344 & 81.17 & -21.09 & 586.0 & 4.0 & NED & 7.70 & 1.20 & 5 & rv & 0.143 & 10.45 &7.44 & 0.04 & 2M \\ 
NGC4826 & 95.61 & 6.13 & 408.0 & 4.0 & NED & 7.48 & 0.69 & 4 & SBF & 0.178 & 9.36 &5.33 & 0.02 & 2M \\ 
NGC3115 & 112.40 & -42.86 & 720.0 & 5.0 & NED & 9.68 & 0.40 & 4 & SBF & 0.205 & 9.87 &5.88 & 0.02 & 2M \\ 
NGC2784 & 136.74 & -56.74 & 691.0 & 35.0 & NED & 9.82 & 1.70 & 4 & SBF & 0.925 & 11.30 &6.32 & 0.02 & 2M \\ 
NGC5068 & 138.28 & -0.21 & 673.0 & 5.0 & NED & 6.90 & 1.20 & 5 & rv & 0.439 & 10.52 &7.55 & 0.05 & 2M \\ 
NGC4945 & 165.18 & -10.21 & 560.0 & 5.0 & NED & 4.10 & 1.20 & 5 & rv & 0.762 & 9.30 &4.48 & 0.02 & 2M \\ 
ESO274G01 & 171.62 & 10.29 & 522.0 & 5.0 & NED & 4.70 & 1.20 & 5 & rv & 1.108 & 11.70 &8.36 & 0.05 & 2M \\ 
Circinus & 183.14 & -6.42 & 449.0 & 10.0 & NED & 2.70 & 1.20 & 5 & rv & 2.000 & 9.61 &4.98 & 0.03 & 2M \\ 
IC5052 & 216.46 & 2.59 & 598.0 & 5.0 & NED & 6.60 & 1.20 & 5 & rv & 0.219 & 11.16 &8.88 & 0.06 & 2M \\ 
NGC55 & 256.25 & -2.36 & 129.0 & 3.0 & NED & 2.10 & 0.50 & 6 & various & 0.057 & 8.84 &6.25 & 0.05 & 2M \\ 
NGC247 & 275.92 & -3.73 & 160.0 & 2.0 & NED & 3.21 & 0.50 & 6 & various & 0.078 & 9.86 &7.43 & 0.06 & 2M \\ 
NGC628 & 314.53 & -5.39 & 657.0 & 1.0 & NED & 7.32 & 1.50 & 7 & stars & 0.301 & 9.95 &6.84 & 0.05 & 2M \\ 
NGC891 & 342.97 & -4.75 & 528.0 & 4.0 & NED & 8.36 & 0.54 & 4 & SBF & 0.280 & 10.81 &5.94 & 0.02 & 2M \\ 
Maffei2 & 359.58 & 0.83 & -17.0 & 5.0 & NED & 2.80 & 1.20 & 8 & TF & 7.190 & 14.27 &5.21 & 0.03 & 2M \\ 
\enddata
\tablerefs{(1), \citet{KSH00}; (2), \citet{MK03}; (3), \citet{DK00}; (4), \citet{T01}; (5), this paper;
(6), \citet{PC88} (modified, see text); (7), \citet{SKT96}; (8), \citet{KKH04}
}
\tablecomments{Observational data on Local Volume galaxies whose distance is not as well known, but which
are among the brightest in the Volume.  TRGB*, (uncertain) tip of the Red Giant Branch; stars, brightest
stars; SBF, surface-brightness fluctuations; rv, radial-velocity (assuming zero peculiar velocity and
applying the kinematic model derived in this paper);
TF, Tully-Fisher relation.  The absorption for Circinus was taken from \citet{FKL77}, all others
from \citet{SFD98} by way of NED.}
\end{deluxetable}
\clearpage
\section{Analysis}

Before we can begin to look at peculiar velocities we need some way of separating them
from background motion.

\subsection{Peculiar Motion}

The motion of a particle $i$ under the gravitational influence of a mass-field $\rho({\bf r}')$ 
is governed by the equation
\begin{equation}
\frac{d {\bf v}_i}{dt} = G \int \frac{\rho ({\bf r'} - {\bf r}_i)}{|{\bf r'} - {\bf r}_i|^3}
d{\bf r}'
\end{equation}
where ${\bf r}_i$ is the location of the particle.
In principle, the (three-dimensional) integral is taken over all space occupied by matter;
in practice, we restrict ourselves to a limited region, and take any
part of the universe outside to be isotropic, and thus to have no effect.

If we separate $\rho$ into a smooth background density $\rho_b$ and a fluctuating part,
and in addition assume that the fluctuations can be adequately modelled by point masses,
the equation of motion becomes
\begin{eqnarray*}
\frac{d {\bf v}_i}{dt} & = & G \int \frac{\rho_b ({\bf r'} - {\bf r}_i)}{|{\bf r'} - {\bf r}_i|^3}
 + G \sum_{j \neq i} \frac{m_j ({\bf r'} - {\bf r}_i)}{|{\bf r'} - {\bf r}_i|^3} \nonumber \\
& = & \frac{d {\bf v}_b}{dt} + \frac{d {\bf v}_f}{dt}
\end{eqnarray*}
so the acceleration splits into that induced by a smooth background ($b$) and that caused
by fluctuations ($f$).  When integrated over time the former will be a smooth Hubble flow, 
the latter the peculiar velocity.  Note that if attention is restricted to a part of the
universe, the background may have a different value than another part; so that,
for instance, the ``Hubble constant'' within 10 Mpc is not necessarily  that of
the universe as a whole.

\subsection{Kinematic Background Solutions}

In practice, we must calculate the background solution from data at hand.
Smooth motion is modelled as a contant average velocity ${\bf v}$ plus an expansion, which
might be anisotropic, represented by a tensor ${\bf H}$.  (Strictly speaking, an
anisotropic expansion reflects peculiar velocity, as one might expect from an external
tidal force; this will be treated below.  It is convenient, however, to calculate it
as part of the background.)  We take as a measure
of goodness of fit of the model to the data
the average square of the difference between
the predicted radial velocity and the observed radial velocity,
\begin{equation}
\sigma^2 = \frac{1}{N} 
\sum_i \sigma_i^2 = \frac{1}{N} \sum_i \left( v_{\rm obs} - \left( {\bf v}_0 \cdot
\hat{\bf r} + \hat{\bf r} \cdot {\bf H} \cdot {\bf r} \right) \right)^2
\label{eq:dispersion}
\end{equation}
The parameters of the model,
${\bf v}_0$ and the six independent components of ${\bf H}$, are
found analytically by taking the derivative of $\sigma^2$ with respect
to each.  Then  setting the resulting expressions
equal to zero gives nine linear equations to be solved for the nine
unknowns, a straightforward if tedious calculation.  An isotropic
solution is determined similarly, using four equations in four unknowns.
(A true maximum-likelihood solution could be constructed by
weighting the various data points with their
observational errors.  As will be seen, in this case the errors are
a minor contributor to scatter about the solutions, and the additional
complication of weighting would not make any significant difference
in the solutions.)

In determining these background solutions, as well as in any calculations
depending on peculiar velocities, only the 149-galaxy data set is used.

It is worth emphasizing that the background solutions are kinematic, and thus
independent of any specific cosmology.  That is, they do not assume a
value for $\rho_b$, or $\Lambda$, or anything else.  The peculiar
velocites derived from them are thus not affected by variations in such
parameters.  

It is important to know just how certain the derived background parameters
are.  For this the error-tensor method of \citet{WH03} is used; details are
set out in that paper.  Briefly, this uses an analytical formula to
measure how much a certain
parameter can be changed before the spread of data points about the solution becomes
significantly greater (using the F-ratio test).  
The numbers quoted correspond to a 68\% chance of the modified
model being significantly worse, and are thus comparable to a standard one-sigma error bar.
There are two obvious ways this approach can fail: first, the distribution of points
around the solution can be non-Gaussian.  In some situations in this paper, for instance
the distribution of errors in synthetic gravity, a Gaussian is very wrong; but not in those
in which I have employed the error tensor.

Second, a change in two or more parameters together may cause less increase in dispersion than
one alone.  This way both are more uncertain than each considered individually might be.  This
is shown by significant off-diagonal terms in the error tensor; for calculations herein, they
were not important.

The model parameters, along with similar results from other efforts, are shown in Table 
(\ref{solutions}).  

\begin{deluxetable}{lcrr}
\tablecaption{Background Solutions \label{solutions}}
\tabletypesize{\footnotesize}
\tablehead{
\colhead{Quantity} & \colhead{Magnitude} & \colhead{L} & \colhead{B} }  
\startdata
 & km s$^{-1}$ or km s$^{-1}$ Mpc$^{-1}$ & Degrees &Degrees  \\
Isotropic & & & \\
V & 336$\pm 52$ & 11$\pm 8$ & 40$\pm 8$  \\
H & 66 $\pm 6$& & \\
$\sigma$ & 79 & & \\
Anisotropic & & & \\
V & 352 & 14 & 48 \\
$H_{xx}$ & 86 $\pm 18$& 119 & 15 \\
$H_{yy}$ & 53 $\pm 8$& 23 & 20 \\
$H_{zz}$ & 39 $\pm$ 20& 242 & 65 \\
$\sigma$ & 72 & & \\
(1) Isotropic (98)& & & \\
V & 290 $\pm 90$ & 10 & 45 \\
H & 64 & & \\
$\sigma$ & 118 & & \\
(1) Anisotropic (98)& & & \\
V & 310 $\pm 90$ & 350 & 49 \\
$H_{xx}$ & 83 $\pm 15$& 127 & 3 \\
$H_{yy}$ & 51 $\pm 14$& 34 & 46 \\
$H_{zz}$ & 32 $\pm 16$ & 320 & 44 \\
$\sigma$ & 103 & & \\
(1) Isotropic (35) & & & \\
V & 350 $\pm 80$& 355 & 44 \\
H & 70 & & \\
$\sigma$ & 89 & & \\
(1) Anisotropic (35) & & & \\
V & 330 $\pm 90$ & 350 & 49 \\
$H_{xx}$ & 138 $\pm 61$& 346 & -65 \\
$H_{yy}$ & 84 $\pm 18$& 104 & -12 \\
$H_{zz}$ & 35 $\pm 16$& 19 & 21 \\
$\sigma$ & 77 & & \\
(2) & & & \\
V & 334 & 22 & 29 \\
(3) Anisotropic & & & \\
V & 325 & 11 & 41 \\
$H_{xx}$ & 82 $\pm 3$ & 132 & 0 \\
$H_{yy}$ & 62 $\pm 3$& 42 & 0 \\
$H_{zz}$ & 48 $\pm 5$& 0 & 90 \\
\enddata
\tablerefs{(1), \citet{WH03}; (2), \citet{YTS77}; (3), \citet{KM01}.}
\tablecomments{Parameters of the local galaxy flow derived in this and other papers:
V, the solar reflex velocity (reciprocal of the average flow with respect to
the Sun); H, effective (isotropic)
Hubble constant; $H_{xx}$ etc., eigenvectors of the anisotropic Hubble flow.  
Listed uncertainties in Hubble tensor components in \citet{KM01} were calculated in a different
way from the other papers.  
The V of \citet{YTS77} was determined using only Local Group galaxies.
The solutions from \citet{WH03} are for the larger (98-galaxy) sample of less reliable
data as well as for a smaller (35-galaxy) sample of better data.}
\end{deluxetable}
\clearpage

The first thing to point out is the velocity dispersion around the models.
Of the 79 and 72 km s$^{-1}$, the listed observational errors contribute
24 km s$^{-1}$, leaving 75 and 66 km s$^{-1}$ of actual motion.  It is worth
emphasizing that these figures are signal, not noise; motions to be explained,
not random errors to be minimized or in which some ``real'' object or correlation
is hiding.

Second, the solar reflex velocity (the reciprocal of the average velocity of the 
galaxies within the Volume) is fairly well-determined, to within less than ten
degrees in direction and ten or twenty km s$^{-1}$.  (The latter is much better
than the assigned error in the latest isotropic calculation; this may be because
the various calculations reuse many of the same data.)

The magnitudes of the components of the Hubble tensor also seem well-determined.
However, the directions vary greatly, so the tensor itself cannot be said to
agree well among the various calculations.  Note the listed (68\%, one-sigma)
uncertainties of the (present paper) eigenvalues: each overlaps the next.  If
90\% uncertainties are used we have $86\pm31$, $53\pm14$ and $40\pm34$
km s$^{-1}$ Mpc$^{-1}$, and all three overlap.  The best that can be said
is that some anisotropy may have been detected at a one-sigma level.  (The
errors listed for \citet{KM01} are formal errors, and appear to be rather
optimistic about the real uncertainty in anisotropic flow.)

How much anisotropy should we expect? \citet{KHK03} performed N-body simulations constrained to
give the same overall structure as found in the local universe (extending beyond
the Local Volume, but with good resolution within it).  In their analogue of the
Supergalactic Plane {\em most} of the peculiar velocities (of something under
100 km s$^{-1}$) were perpendicular
to that feature; in other words, the major component of peculiar velocities
reflected infall into the Plane.  In our case, at most 7 or 8 km s$^{-1}$
(the difference between isotropic and anisotropic solutions) out of
75 km s$^{-1}$ can be attributed to SGP infall, about one-tenth.

This is an important result.  Whether from dynamical or kinematical arguments,
we expect to see a great deal of infall into the largest structure in the Volume,
and it's not there.

With the failure of the tensor solution to show much beyond the isotropic
solution, we will take the latter as the basis for further calculations.
Table (\ref{data3})
shows the absolute magnitudes (in $B$ and $K$), with uncertainties, and the
peculiar radial velocities for each of the 170 galaxies in the total data set,
based on that background.

\begin{deluxetable}{lrrrrl}
\tablecaption{Derived Local Volume Data \label{data3}}
\tablehead{
\colhead{Name} & \colhead{$M_B$} & \colhead{$\Delta M_B$} & \colhead{$M_K$} & \colhead{$\Delta M_K$}
& \colhead{Peculiar RV}}
\startdata
NGC6946 & -21.03 & 0.60 & -23.79 & 0.45 & -75 \\ 
KK98-35 & -12.64 & 0.46 & -10.00 & 0.22 & -30 \\ 
IC342 & -20.89 & 0.44 & -23.02 & 0.18 & 59 \\ 
NGC1569 & -17.61 & 0.46 & -18.59 & 0.22 & -12 \\ 
UGCA105 & -14.94 & 0.46 & -10.00 & 0.22 & 100 \\ 
KK98-44 & -12.45 & 0.40 & -10.00 & 0.00 & 75 \\ 
NGC1560 & -16.34 & 0.46 & -18.76 & 0.24 & -28 \\ 
KK98-41 & -14.08 & 0.48 & -10.00 & 0.26 & -67 \\ 
KKH34 & -12.29 & 0.43 & -10.00 & 0.17 & 23 \\ 
NGC6789 & -14.34 & 0.46 & -15.56 & 0.24 & -77 \\ 
NGC2366 & -16.25 & 0.49 & -16.90 & 0.31 & 70 \\ 
NGC2403 & -18.74 & 0.44 & -21.31 & 0.18 & 85 \\ 
NGC6503 & -17.84 & 0.51 & -21.31 & 0.32 & -11 \\ 
DDO50 & -16.64 & 0.41 & -18.80 & 0.11 & 117 \\ 
KDG52 & -11.34 & 0.43 & -10.00 & 0.16 & 58 \\ 
UGC4483 & -12.61 & 0.41 & -10.00 & 0.09 & 111 \\ 
DDO53 & -13.44 & 0.43 & -10.00 & 0.15 & -61 \\ 
UGC3755 & -14.82 & 0.42 & -10.00 & 0.11 & -133 \\ 
VIIZw403 & -13.88 & 0.40 & -15.62 & 0.11 & -176 \\ 
HoI & -15.13 & 0.48 & -10.00 & 0.26 & 64 \\ 
M82 & -19.33 & 0.45 & -23.29 & 0.21 & 113 \\ 
BK3N & -11.45 & 0.42 & -10.00 & 0.13 & -160 \\ 
M81 & -20.21 & 0.41 & -23.92 & 0.08 & -105 \\ 
NGC2976 & -17.24 & 0.46 & -20.23 & 0.23 & -74 \\ 
KK98-81 & -12.49 & 0.43 & -10.00 & 0.15 & -211 \\ 
NGC3077 & -17.61 & 0.45 & -20.63 & 0.21 & -78 \\ 
Garland & -10.00 & 0.47 & -10.00 & 0.24 & -40 \\ 
DDO82 & -14.73 & 0.46 & -10.00 & 0.22 & 88 \\ 
KK98-85 & -10.08 & 0.43 & -10.00 & 0.15 & -105 \\ 
DDO71 & -10.13 & 0.43 & -10.00 & 0.15 & -206 \\ 
IC2574 & -17.38 & 0.46 & -17.30 & 0.25 & -47 \\ 
Draco & -13.62 & 0.40 & -10.00 & 0.03 & -85 \\ 
KDG73 & -13.09 & 0.42 & -10.00 & 0.13 & -216 \\ 
DDO78 & -12.15 & 0.43 & -10.00 & 0.15 & -34 \\ 
DDO47 & -15.12 & 0.47 & -10.00 & 0.25 & -164 \\ 
KK98-65 & -12.81 & 0.44 & -10.00 & 0.18 & 13 \\ 
NGC4236 & -18.25 & 0.46 & -19.23 & 0.22 & -115 \\ 
UMi & -7.64 & 0.42 & -10.00 & 0.11 & -34 \\ 
DDO165 & -15.60 & 0.44 & -10.00 & 0.19 & -82 \\ 
UGC4115 & -13.62 & 0.46 & -10.00 & 0.23 & -139 \\ 
NGC2683 & -19.36 & 0.58 & -23.49 & 0.43 & -224 \\ 
KKR25 & -9.38 & 0.42 & -10.00 & 0.14 & -41 \\ 
NGC5204 & -16.66 & 0.47 & -18.83 & 0.25 & 45 \\ 
NGC3738 & -16.37 & 0.47 & -10.00 & 0.24 & 2 \\ 
NGC5585 & -18.56 & 0.58 & -20.20 & 0.43 & -100 \\ 
UGC8508 & -12.71 & 0.42 & -10.00 & 0.13 & 28 \\ 
M101 & -20.90 & 0.41 & -23.66 & 0.11 & -60 \\ 
UGC7298 & -13.22 & 0.43 & -10.00 & 0.17 & -11 \\ 
UGC6541 & -10.00 & 0.48 & -10.00 & 0.26 & 58 \\ 
NGC3741 & -13.22 & 0.46 & -10.00 & 0.24 & 73 \\ 
M106 & -20.34 & 0.41 & -23.91 & 0.08 & 26 \\ 
LeoA & -11.52 & 0.44 & -10.00 & 0.18 & -84 \\ 
KK98-109 & -10.85 & 0.43 & -10.00 & 0.16 & -34 \\ 
NGC5195 & -19.13 & 0.48 & -23.17 & 0.27 & 62 \\ 
NGC5194 & -20.61 & 0.48 & -23.93 & 0.27 & 59 \\ 
DDO167 & -11.15 & 0.47 & -10.00 & 0.24 & -25 \\ 
DDO168 & -15.55 & 0.47 & -10.00 & 0.25 & -7 \\ 
NGC4449 & -18.21 & 0.48 & -20.87 & 0.26 & -15 \\ 
DDO125 & -14.25 & 0.41 & -10.00 & 0.10 & 82 \\ 
NGC2903 & -20.20 & 0.59 & -23.71 & 0.44 & -133 \\ 
DDO190 & -14.08 & 0.42 & -10.00 & 0.12 & 80 \\ 
DDO99 & -13.82 & 0.44 & -10.00 & 0.17 & 80 \\ 
M94 & -19.43 & 0.49 & -23.24 & 0.28 & 51 \\ 
NGC4244 & -17.47 & 0.46 & -20.54 & 0.23 & -36 \\ 
UGCA290 & -10.00 & 0.42 & -10.00 & 0.12 & 41 \\ 
DDO181 & -12.72 & 0.45 & -10.00 & 0.21 & 77 \\ 
IC3687 & -14.68 & 0.46 & -10.00 & 0.23 & 85 \\ 
DDO126 & -14.30 & 0.47 & -10.00 & 0.25 & -85 \\ 
NGC4214 & -17.11 & 0.42 & -19.34 & 0.14 & 113 \\ 
DDO113 & -11.97 & 0.41 & -10.00 & 0.11 & 102 \\ 
IC4182 & -15.34 & 0.40 & -10.00 & 0.06 & 62 \\ 
UGC7605 & -13.30 & 0.48 & -10.00 & 0.26 & 28 \\ 
NGC3344 & -19.13 & 0.52 & -22.00 & 0.34 & 0 \\ 
NGC4395 & -17.68 & 0.43 & -18.27 & 0.18 & 23 \\ 
UGC8833 & -11.07 & 0.42 & -10.00 & 0.14 & 78 \\ 
LeoI & -11.05 & 0.41 & -10.00 & 0.08 & 111 \\ 
M95 & -19.44 & 0.41 & -23.18 & 0.10 & 10 \\ 
M96 & -19.97 & 0.40 & -23.65 & 0.06 & 96 \\ 
SextansB & -13.91 & 0.41 & -10.00 & 0.07 & 22 \\ 
NGC4826 & -20.19 & 0.45 & -24.04 & 0.20 & -132 \\ 
M66 & -20.35 & 0.41 & -23.98 & 0.09 & -28 \\ 
DDO187 & -12.71 & 0.44 & -10.00 & 0.17 & 3 \\ 
GR8 & -12.16 & 0.47 & -10.00 & 0.25 & -18 \\ 
Sextans & -8.10 & 0.40 & -10.00 & 0.05 & -6 \\ 
SextansA & -13.93 & 0.41 & -10.00 & 0.07 & 1 \\ 
NGC3115 & -20.26 & 0.41 & -24.05 & 0.09 & -163 \\ 
KKH86 & -10.40 & 0.42 & -10.00 & 0.13 & 18 \\ 
NGC2784 & -19.59 & 0.55 & -23.64 & 0.38 & -251 \\ 
NGC3109 & -15.50 & 0.40 & -16.32 & 0.09 & 9 \\ 
NGC5068 & -19.11 & 0.55 & -21.65 & 0.38 & 0 \\ 
Antlia & -9.75 & 0.43 & -10.00 & 0.16 & -46 \\ 
NGC3621 & -19.21 & 0.41 & -22.44 & 0.11 & -5 \\ 
PGC48111 & -13.65 & 0.44 & -10.00 & 0.19 & 53 \\ 
KK98-112 & -12.31 & 0.44 & -10.00 & 0.18 & -2 \\ 
IC4316 & -13.49 & 0.46 & -10.00 & 0.23 & 153 \\ 
KK98-208 & -14.24 & 0.44 & -10.00 & 0.19 & -163 \\ 
M83 & -20.24 & 0.43 & -23.53 & 0.16 & -1 \\ 
KK98-200 & -11.96 & 0.43 & -10.00 & 0.15 & -67 \\ 
NGC5264 & -15.90 & 0.48 & -17.74 & 0.27 & -53 \\ 
NGC5253 & -16.93 & 0.42 & -19.31 & 0.15 & -49 \\ 
PGC39032 & -12.75 & 0.43 & -10.00 & 0.17 & 104 \\ 
NGC5102 & -17.54 & 0.47 & -20.74 & 0.25 & -19 \\ 
PGC47171 & -15.42 & 0.47 & -10.00 & 0.25 & -1 \\ 
NGC5128 & -20.53 & 0.43 & -23.93 & 0.16 & 28 \\ 
PGC48738 & -14.04 & 0.47 & -10.00 & 0.25 & 58 \\ 
NGC5408 & -16.51 & 0.44 & -17.02 & 0.22 & -55 \\ 
NGC4945 & -19.53 & 0.75 & -23.58 & 0.64 & 0 \\ 
PGC51659 & -11.83 & 0.45 & -10.00 & 0.20 & -81 \\ 
ESO274G01 & -17.77 & 0.68 & -20.00 & 0.56 & 0 \\ 
Circinus & -19.55 & 1.04 & -22.18 & 0.97 & 0 \\ 
MilkyWay & -20.50 & 0.48 & -23.70 & 0.57 & -49 \\ 
IC3104 & -14.85 & 0.44 & -13.18 & 0.39 & -3 \\ 
NGC2915 & -15.82 & 0.48 & -18.06 & 0.26 & -78 \\ 
PGC20125 & -14.76 & 0.46 & -10.00 & 0.22 & -67 \\ 
Carina & -9.24 & 0.47 & -10.00 & 0.24 & -64 \\ 
LMC & -17.92 & 0.40 & -10.00 & 0.00 & 8 \\ 
IC5052 & -18.16 & 0.56 & -20.22 & 0.40 & 0 \\ 
SagDIG & -10.13 & 0.41 & -10.00 & 0.10 & -60 \\ 
SMC & -16.35 & 0.40 & -10.00 & 0.00 & -44 \\ 
PGC19337 & -14.78 & 0.46 & -10.00 & 0.23 & -14 \\ 
Tucana & -9.13 & 0.43 & -10.00 & 0.15 & -14 \\ 
NGC1313 & -19.35 & 0.40 & -20.51 & 0.08 & -15 \\ 
NGC6822 & -15.12 & 0.41 & -16.69 & 0.10 & 20 \\ 
NGC1705 & -15.81 & 0.48 & -18.02 & 0.27 & 58 \\ 
IC5152 & -15.42 & 0.46 & -17.10 & 0.23 & -63 \\ 
PGC09962 & -15.21 & 0.46 & -14.77 & 0.29 & 20 \\ 
KK98-54 & -13.17 & 0.48 & -10.00 & 0.26 & -61 \\ 
DDO210 & -11.09 & 0.41 & -10.00 & 0.07 & -59 \\ 
Phoenix & -10.01 & 0.40 & -10.00 & 0.05 & -136 \\ 
PGC01641 & -10.84 & 0.41 & -10.00 & 0.10 & -50 \\ 
ESO245-05 & -15.60 & 0.46 & -10.00 & 0.22 & 16 \\ 
NGC55 & -17.83 & 0.65 & -20.36 & 0.52 & -31 \\ 
NGC625 & -16.31 & 0.42 & -18.86 & 0.13 & 68 \\ 
UGCA438 & -12.88 & 0.41 & -10.00 & 0.10 & -46 \\ 
NGC300 & -17.63 & 0.41 & -20.15 & 0.10 & -21 \\ 
UGCA442 & -14.62 & 0.48 & -15.38 & 0.43 & 17 \\ 
NGC7793 & -18.06 & 0.46 & -21.10 & 0.24 & -5 \\ 
Sculptor & -9.49 & 0.42 & -10.00 & 0.11 & 90 \\ 
Fornax & -11.47 & 0.43 & -10.00 & 0.14 & -39 \\ 
NGC253 & -20.02 & 0.45 & -24.21 & 0.20 & 20 \\ 
IC1574 & -13.43 & 0.47 & -10.00 & 0.26 & 94 \\ 
DDO6 & -12.50 & 0.43 & -10.00 & 0.16 & 140 \\ 
NGC247 & -17.75 & 0.52 & -20.10 & 0.34 & 10 \\ 
WLM & -13.88 & 0.41 & -10.00 & 0.10 & -62 \\ 
IC1613 & -14.54 & 0.40 & -11.55 & 0.12 & -122 \\ 
DDO216 & -11.48 & 0.49 & -10.00 & 0.29 & 26 \\ 
UGC685 & -14.42 & 0.42 & -10.00 & 0.14 & 59 \\ 
NGC628 & -19.67 & 0.60 & -22.48 & 0.45 & 370 \\ 
AndVI & -10.54 & 0.41 & -10.00 & 0.08 & -124 \\ 
LGS3 & -9.94 & 0.41 & -10.00 & 0.07 & -82 \\ 
KK98-16 & -12.38 & 0.47 & -10.00 & 0.24 & 114 \\ 
M33 & -18.44 & 0.40 & -20.43 & 0.07 & 10 \\ 
KK98-17 & -11.41 & 0.44 & -10.00 & 0.19 & 64 \\ 
AndII & -10.87 & 0.40 & -10.00 & 0.03 & 28 \\ 
AndIII & -9.65 & 0.45 & -10.00 & 0.20 & -113 \\ 
NGC404 & -16.57 & 0.42 & -18.96 & 0.15 & 11 \\ 
KKH98 & -10.78 & 0.41 & -10.00 & 0.11 & 15 \\ 
AndI & -11.04 & 0.41 & -10.00 & 0.07 & -132 \\ 
NGC925 & -19.44 & 0.40 & -21.06 & 0.05 & 169 \\ 
M31 & -20.28 & 0.40 & -23.40 & 0.06 & -56 \\ 
NGC205 & -15.65 & 0.45 & -18.71 & 0.21 & 6 \\ 
KKH18 & -12.39 & 0.47 & -10.00 & 0.24 & 112 \\ 
NGC891 & -19.08 & 0.42 & -23.67 & 0.14 & 219 \\ 
NGC185 & -14.65 & 0.45 & -17.40 & 0.22 & 60 \\ 
NGC147 & -14.68 & 0.49 & -17.20 & 0.29 & 62 \\ 
KKH96 & -11.73 & 0.41 & -10.00 & 0.11 & -33 \\ 
KKH5 & -12.27 & 0.43 & -10.00 & 0.17 & 74 \\ 
IC10 & -18.89 & 0.48 & -18.09 & 0.26 & -80 \\ 
MaffeiI & -20.04 & 0.45 & -22.71 & 0.22 & 129 \\ 
Maffei2 & -20.16 & 1.01 & -22.02 & 0.93 & 57 \\ 
\enddata
\tablecomments{Derived absolute magnitudes of Local Volume galaxies in $B$ and $K$,
with uncertainties taking into account distance and photometric errors; along
with the peculiar radial velocities relative to the best-fit isotropic
expansion.  Galaxies for which no photometric data were available were
arbitrarily assigned an absolute magnitude of -10.  Their listed uncertainties,
which are not used in any calculations, show the effect of distance uncertainties
alone.}
\end{deluxetable}
\clearpage

As will become clearer below, it is important to know how far to
trust the derived peculiar velocities.  They are uncertain because
of the uncertainties in the model parameters, and because of uncertainties
in distance (uncertainties in angle having negligible effect).
For situations in which a derived quantity $Q$ is a function of
various independent parameters $p_i$, each with an uncertainty $\delta p_i$, I have used the formula
\begin{equation}
\left( \delta Q \right)^2 = \sum_i \left( \frac{\partial Q}{\partial p_i}  
\delta p_i \right)^2
\label{error1}
\end{equation}
which is sufficiently accurate as long as the errors are not large fractions of the values.
For the uncertainties in peculiar velocities due to model parameter uncertainties and
distance errors ($\delta D$) this technique gives
\begin{equation}
\left( \delta pv_i \right)^2 = x_i^2 \delta V_x^2 + y_i^2 \delta V_y^2 + z_i^2 \delta V_z^2
+ D_i^2 \delta H^2 + H^2 \delta D^2
\end{equation}
And for the uncertainties in absolute magnitude $M$
due to errors in apparent magnitude $m$ and distance $D$,
\begin{equation}
\delta M^2 = \delta m^2 + \left( \frac{5}{\ln 10} \frac{\delta D}{D} \right)^2
\end{equation}
All $B$ photometry is assigned a $\delta m$ of 0.4 magnitude, while $K$ uncertainties
are taken from the source.  The results are listed in Table (\ref{data3}).  The 
uncertainties in absolute magnitude for those galaxies without photometry reflects
errors in distances alone, and are left in the table to give an idea of the
relative importance of distance and photometry uncertainties.

\subsection{Peculiar Velocity Dispersions: Bright Galaxies}

We now turn to smaller-scale motions, those which can be connected with individual
galaxies or galaxy groups.  For this the dynamical youth of the Local Volume is invoked:
given the peculiar velocity dispertion of 70-80 km s$^{-1}$, we expect that satellite
galaxies around a mass concentration to be mostly still infalling.

To look for this, we pick out the brightest galaxies, those with $M_B$ of -20
or brighter.  (While one can argue about the role of starbursts, for example,
in skewing the mass-to-light ratio, it is certain that $M_B$ of -20 {\em does}
select the very biggest concentrations of luminous matter in the Local Volume.  They
are also the brightest in $K$.)  We turn our attention to those galaxies between
0.5 and 1.5 Mpc away from the bright ones, to isolate a sample which should feel their gravity
but which probably haven't crossed from one side to the other (near to far, or vice
versa).  Finally, we choose
only those satellites in front of or behind the giants (with the relative direction within
60 degrees of the line of sight), so that any motion due to infall shows up well in
radial velocity.

To see the order of magnitude of the effect we are looking for, consider a galaxy
of $10^{12}$ solar masses, acting at a distance of 1 Mpc over 10-13 billion years.
That results in a change in speed of 44-67 km s$^{-1}$; so we look for something
in the tens, but not hundreds, of km s$^{-1}$.  The average of satellites behind
the giant galaxy should show a blueshift (negative), and those before a redshift;
the difference should be on the order of 100 km s$^{-1}$.  There should be some
noise from those satellites having made a close approach, but given the observed
peculiar velocities we do not expect very many of them.
There are not enough satellite galaxies around most of the bright ones to give a clear
picture, so the data from all groups outside the Local Group
are averaged together in Table (\ref{cherry}).  There being no satellites
``before'' the Milky Way, and only one for M31, the numbers listed
for those giants are all ``behind''.

\begin{deluxetable}{lcccc}
\tablecaption{Satellite Galaxy Motions \label{cherry}}
\tablehead{
\colhead{Quantity} & \colhead{Before} & \colhead{Behind} & \colhead{M31} & \colhead{Milky Way}}
\startdata
 & km s$^{-1}$ & km s$^{-1}$ & km s$^{-1}$ & km s$^{-1}$\\
Mean & 10$\pm 27$ & -39$\pm 3$ & -74$\pm 85$ & -73$\pm 37$ \\
rms dispersion & 61 & 107 & 76 & 66 \\
Number of satellites & 7 & 17 & 10 & 24 \\
%Average negative (Number) & -31 (4) & -95 (12) & -88 (9) & -86 (21) \\
%Average positive (Number) & 65 (3) & 95 (5) & 54 (1) & 18 (3) \\
\enddata
\tablecomments{Satellite galaxy peculiar radial velocities relative to bright ($M_B < -20$)
galaxies, from the combined Local Volume sample.  Satellites are between 0.5 and 1.5 Mpc
from the giant, and sorted into those in front and those behind; those at a similar
distance are not included.  The satellites of M31 and the Milky Way are numerous enough
to show separately; all these are ``behind'', in the sense that infall would show as a
negative difference in peculiar velocity; except for Carina in the case of M31, which
is not included here.}
\end{deluxetable}
\clearpage

In half of the cases there is no significant sign of infall.  In another (the Volume average
``behind'') the rms dispersion is much larger than the average, so infall contributes only
a minor part of the peculiar velocity.  There is a fairly clear sign of infall into the
Milky Way.  But if an effect is seen in only half (or less) of the places it should be
it can hardly be accepted as a general feature of the situation, and
there is the strong possibility that the positive signs are due to something else.  

The method of computing the uncertainties here is that used for
averages of data points, each of which has an assigned uncertainty $\sigma_i$;
which uncertainties are uncorrelated with each other.
I have used the following formula for the uncertainty of the average, $\sigma_T$:
\begin{equation}
\frac{1}{\sigma^2_T} = \sum_{i} \frac{1}{\sigma^2_i}.
\end{equation}
Now, the uncertainties among satellite galaxy peculiar velocities {\em are} correlated,
since the largest part comes from parameters in the model and changing a parameter for one
will change the parameter for all.  That would require, formally, that uncertainties be
added, resulting in a much larger overall uncertainty.  On the other hand, for satellites about the
same giant any change in parameters will affect them all in much the same way, causing {\em less}
overall uncertainty.  Rather than spend more time on this rather marginal calculation
attempting to sort out uncertainties, however, I will note that the rms dispersions alone
show that infall is not a strong signal, even if present; and treat the effects of changes in
the model in more detail in a later section.

These results are surprising not least in the fact that \citet{KSD03} and \citet{KMS03}
found frontside infall but no backside infall in Local Volume groups.  Most of this I
will trace to differences in the background model, of which (again) I postpone a detailed
discussion for a later section.  Part could be due to spurious signals in poor or limited
data.  As examples of the latter there is the peculiar velocity versus Supergalactic Z
plot, Figure 6 in \citet{KM96}, which shows a clear signal of infall into the Supergalactic Plane,
a signal which disappears with more and better data; and the lopsided histogram of
peculiar velocities of \citet{WH03}, which led to the conclusion that many nearby galaxies of
high radial velocity were being missed--a conclusion which also disappears with the
present collection of data.

\begin{figure}
\plottwo{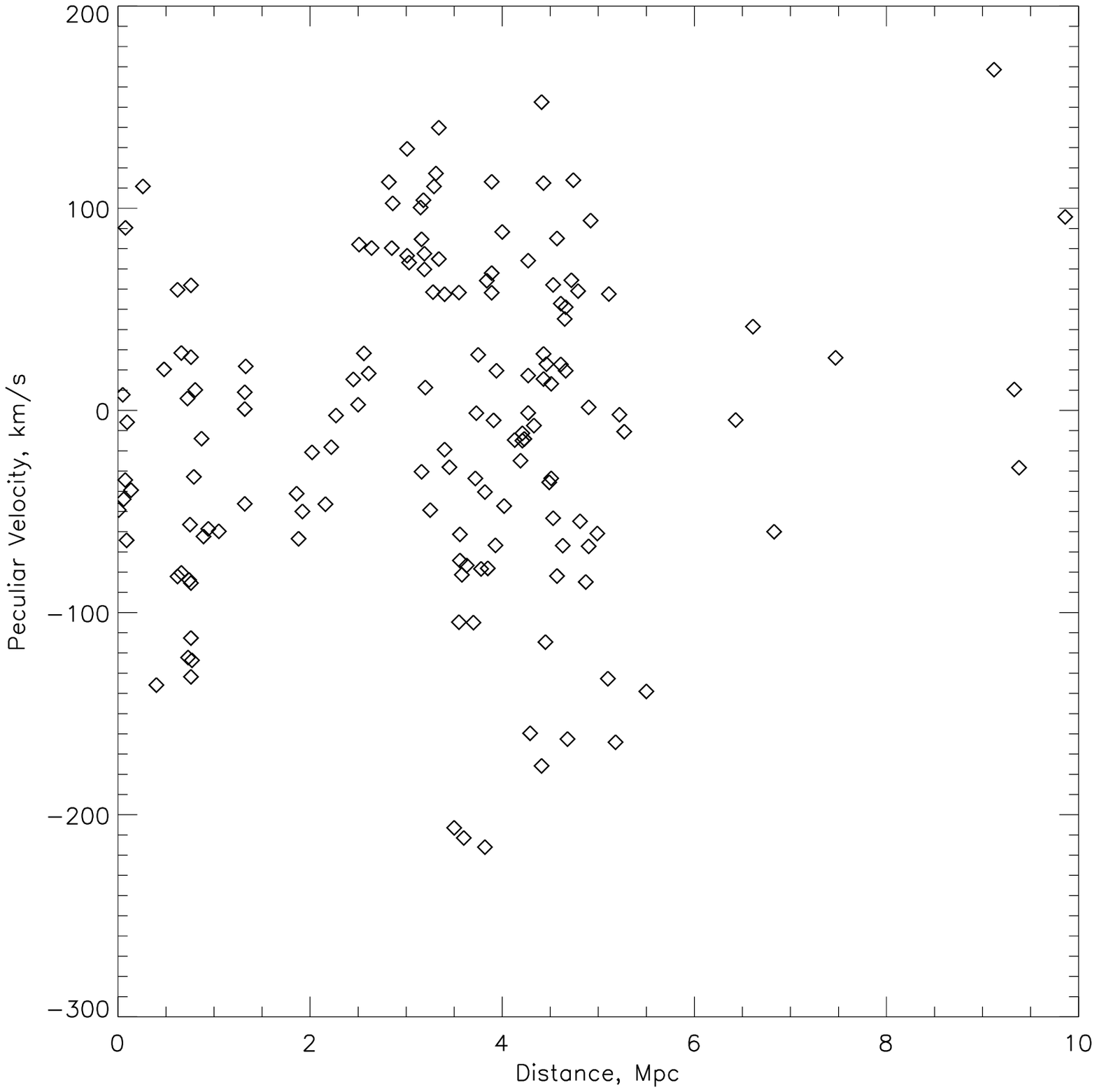}{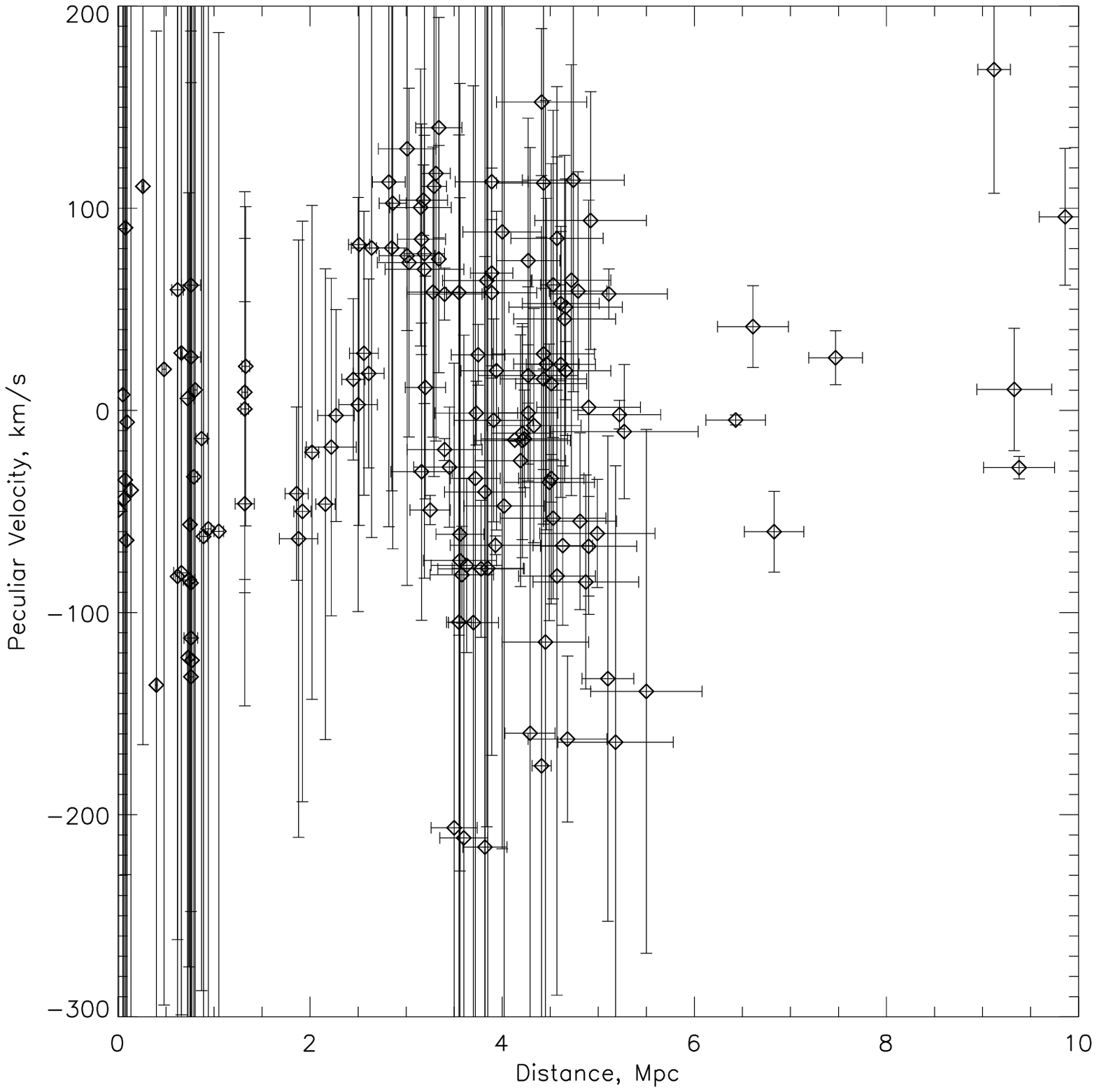}
\caption{Peculiar velocity as a function of distance (for the isotropic model).  Left,
without error bars to show apparent structure; right, with error bars.}
\label{distance}
\end{figure}

Additionally, consider Figure (\ref{distance}).  It is tempting to see in it
an increase in peculiar velocity dispersion around 4 Mpc, due to the M81 and Centaurus A groups.

In fact there is no significant increase in peculiar velocity dispersion.  Within 1.5
Mpc the dispersion is 70 km s$^{-1}$; from 3 to 5 Mpc, 84 km s$^{-1}$, which
is statistically indistinguishable (with this number of data points).  (Note that
both are indistinguishable from the peculiar velocity dispersion of the Volume as
a whole.)  There are more
points on the plot at the larger distance, hence the distribution is more fully shown
including the wings; but it is essentially the same distribution.
Similarly, there is a diagonal line of points between about 1.5 and 2.5 Mpc, which
could easily be taken as a sign of front side infall into Centaurus or M81.  But this is made
up of galaxies in at least four widely-separated parts of the sky, so attributing
a coordinated motion to M81 or Centaurus is questionable; and anyway the structure disappears when
error bars are introduced (as in the right panel of Figure (\ref{distance})).

So it is at least plausible that apparently clear signs of infall seen in other
investigations are mistaken.  We now turn to other ways of relating peculiar
motions to luminous matter.

Perhaps the assumption of dynamical youth implicit in the idea of infall is not right,
at least close to the giants.  If these regions are dynamically old, that is if the
groups are virialized, then there should be a much larger velocity dispersion close
to the giant compared with farther out.  (It should also be true more generally that
satellite galaxies deeper inside a potential well show higher speeds, as potential
energy is converted to kinetic.)  To look for this we
choose two zones, an inner zone out to 1 Mpc from the giant and an outer zone from 1 to 2
Mpc away.  The resulating velocity dispersions are shown in Table
(\ref{apple}).  Included is an F-ratio calculation, which shows the probability of
the outer sample being drawn from a significantly different (smaller) parent
distribution.

\begin{deluxetable}{lccc}
\tablecaption{Satellite Galaxy Velocity Dispersions \label{apple}}
\tablehead{
\colhead{Subsample} & \colhead{Inner} & \colhead{Outer} & \colhead{Significance}}
\startdata
 & km s$^{-1}$ & km s$^{-1}$ &\\
non-Local Group & 114 & 108 & 0.64\\
M31 & 83 & 64 & 0.83\\
Milky Way & 94 & 69 & 0.82\\
\enddata
\tablecomments{Velocity dispersion of satellites around bright galaxies in the Local
Volume (rms difference from the giant), for various subsamples.  Inner galaxies are
less than 1 Mpc from the giant, outer satellites between 1 and 2 Mpc.  The last column
shows the significance level at which the hypothesis of equal inner and outer dispersions
is rejected by the F-ratio test.}
\end{deluxetable}
\clearpage

There does seem to be a smaller velocity dispersion farther from the giants, but it's
not even of one-sigma significance outside the Local Group, and not much better within.
(These should probably be taken as upper limits on the significance, since the satellite
galaxy dispersions are not really Gaussian.)  How much should we see?

\citet{PVK03} examined isolated giant and satellite galaxy pairs in the SDSS data, calculating the
line of sight velocity dispersion for 3000 satellites at various distances from the
primary.  For an
L* galaxy (which is comparable to those examined above), the dispersion fell from
120 km s$^{-1}$ at 20 kpc to 60 km s$^{-1}$ at 350 kpc.   At this rate the giant-satellite
dispersion would be a minor part of the dispersion as observed in the Local Volume 
at megaparsec scales.  In all,
it is unlikely that we have observed any influence of large galaxies on the velocity
dispersion of satellites between 1 and 2 Mpc and much more likely that what has 
been observed is due to some other effect.
(If we separate the
sample from beyond the Local Group into ``behind and before'' and ``beside'', the
inner and outer ``beside'' samples have the same dispersion to within one km s$^{-1}$;
which, at 120 km s$^{-1}$, is {\em larger} than the ``before and behind'' inner
sample.  But selecting data in this way can have a pernicious effect on statistics,
and at the very least increases the chance that small-number fluctuations will 
give misleading results.)

\subsection{The Peculiar Gravity Field}

So far we have dealt with only a subset of our data, those galaxies near the very
brightest.  If we wish to make truly general comparisons of peculiar velocities
with luminous matter, as well as make use of a great deal of data painstakingly
gathered, we should find a way to look at all the galaxies in the sample.

For this we again make use of the dynamical youth of the Local Volume, which allows
us to compare apples with oranges.  Assume that
peculiar velocities are produced by fluctuations in the gravitational field which
can be identified with galaxies.  If the fluctuations are linear (as they are on
much larger scales than we are looking at),
the local gravitational acceleration and the local peculiar velocity
should be proportional: plotting one against the other gives a straight line.
In the mildly nonlinear regime the line will become an S-shaped curve, as the effects
of mass concentrations reinforce themselves.  As some objects complete infall and
pass to the other side of larger masses, the ends of the S will fray; but as long as
the system is not dynamically relaxed there will be some discernable relation between
the gravity field and the peculiar velocity field: apples and oranges.

If possible, we want to avoid the worst inaccuracies in this kind of calculation.  
Galaxies which are at similar distances
and close to each other in the sky will have large calculated gravitational accelerations,
but distance errors as well as motions mean that they may actually be in front when
the calculation has them behind, and vice versa.  To reduce this problem we adopt a
smoothing parameter $A$, which is set to the distance an object with a speed of
78 km s$^{-1}$ will go in the age of the universe, about 1.2 Mpc (setting this
parameter to anything between 1 and 2 Mpc makes little difference).

We proceed, then, by finding the smoothed gravitational acceleration at each of our
galaxies, using the luminosity weights $w_B$ or $w_K$, and the smoothing
length $A$; then find the component in the
radial direction ${\bf \hat{r}}$:
\begin{equation}
g_i = {\bf g}_i \cdot {\bf \hat{r}}_i = G \sum_{j \neq i} w_j 
\frac{{\bf r}_j - {\bf r}_i}{(({\bf r}_j - {\bf r}_i)^2 + A^2)^{3/2}} 
\cdot \hat{\bf {r}}_i
\end{equation}
This we shall call the synthetic gravity, or sometimes the synthetic peculiar gravity.
There is a normalization constant implied in the weights $w_j$, which depends on details of
dynamical model one prefers.  For the moment we leave the synthetic gravity
in arbitrary numbers.  

When calculated this way, the radial components of the gravitational force become steadily
more negative as we proceed to the outskirts of the Volume.  This is as expected.  The average
mass density of the Volume will exert a force inward, increasing linearly with distance
from the center.  Since we are interested in deviations from the average, however, we fit a linear
force to the calculation above and subtract it.

Of course, the uncertainty in the synthetic gravity is important.
These will be due to errors in mass weight $w_j$ (proportional to
$\delta M$ above) and distances $r_j$ and $r_i$, and using the formulation in Equation(\ref{error1})
we have
\begin{eqnarray*}
\delta g_i^2 & = & \sum_{j \neq i} \left[ \left( \frac{r_j \hat{\bf r}_j \cdot \hat{\bf r}_i - r_i}
{\left( ({\bf r}_j-{\bf r}_i)^2 + A^2 \right)^{3/2}} \delta w_j \right)^2 \right. \\
& +& \left. \left(\frac{w_j \hat{\bf r}_j \cdot \hat{\bf r}_i}{\left( ({\bf r}_j-{\bf r}_i)^2 + A^2 \right)^{3/2}}
-3 \frac{w_j (r_j \hat{\bf r}_j \cdot \hat{\bf r}_i-r_i)(r_j-r_i \hat{\bf r}_j \cdot \hat{\bf r}_i)}
{\left( ({\bf r}_j-{\bf r}_i)^2 + A^2 \right)^{5/2}} \right)^2 \delta r_j^2 \right] \\
 & + & \left[ \sum_{j \neq i} \left( \frac{-w_j}{\left( ({\bf r}_j-{\bf r}_i)^2 + A^2 \right)^{3/2}}
+3 w_j \frac{(r_j \hat{\bf r}_j \cdot \hat{\bf r}_i -r_i)^2}
{\left( ({\bf r}_j-{\bf r}_i)^2 + A^2 \right)^{5/2}} \right) \right]^2 \delta r_i^2 
\end{eqnarray*}
where $\hat{\bf r}_j$ is the unit vector in the ${\bf r}_j$ direction, and similarly.

\begin{figure}
\plottwo{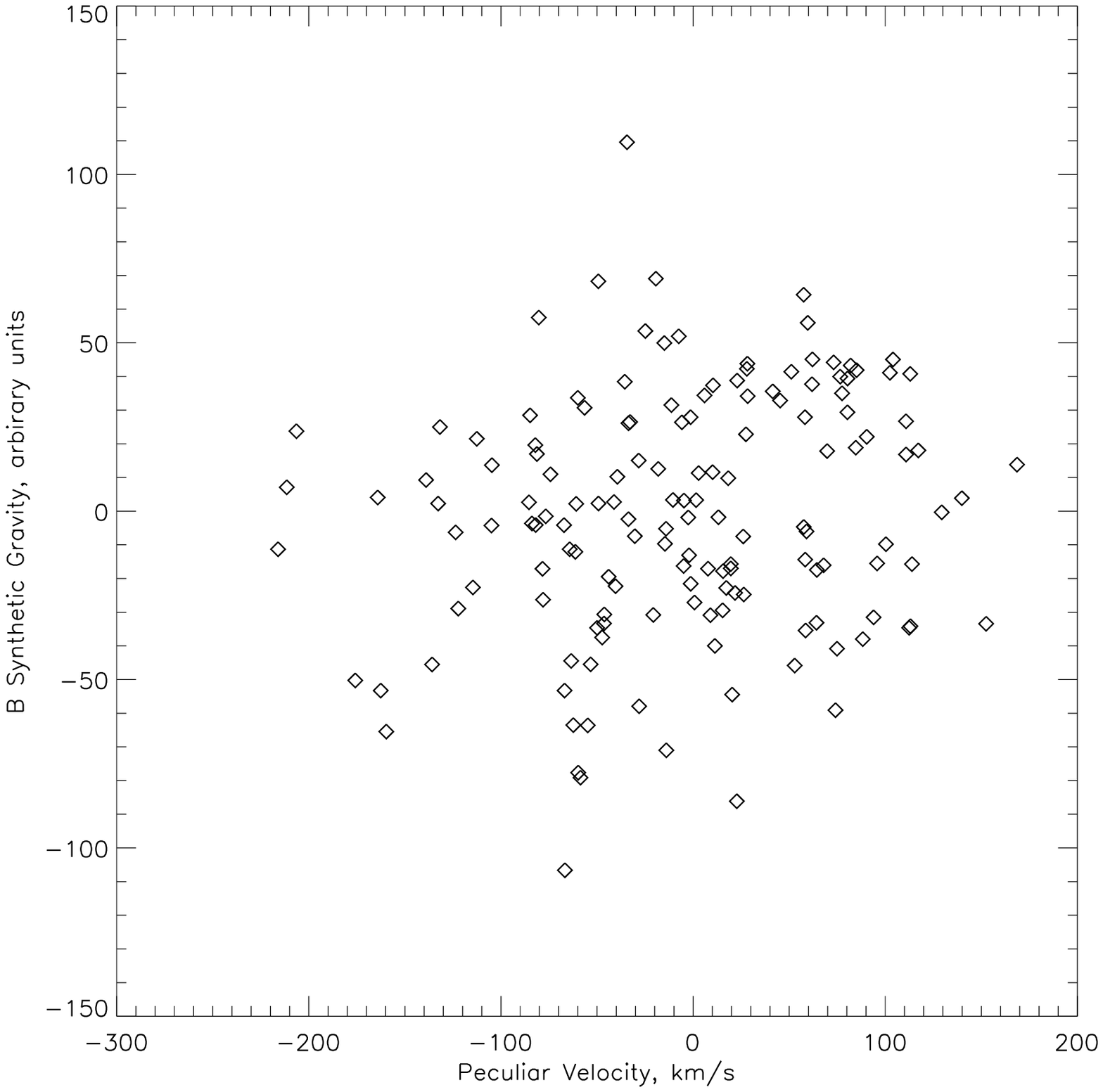}{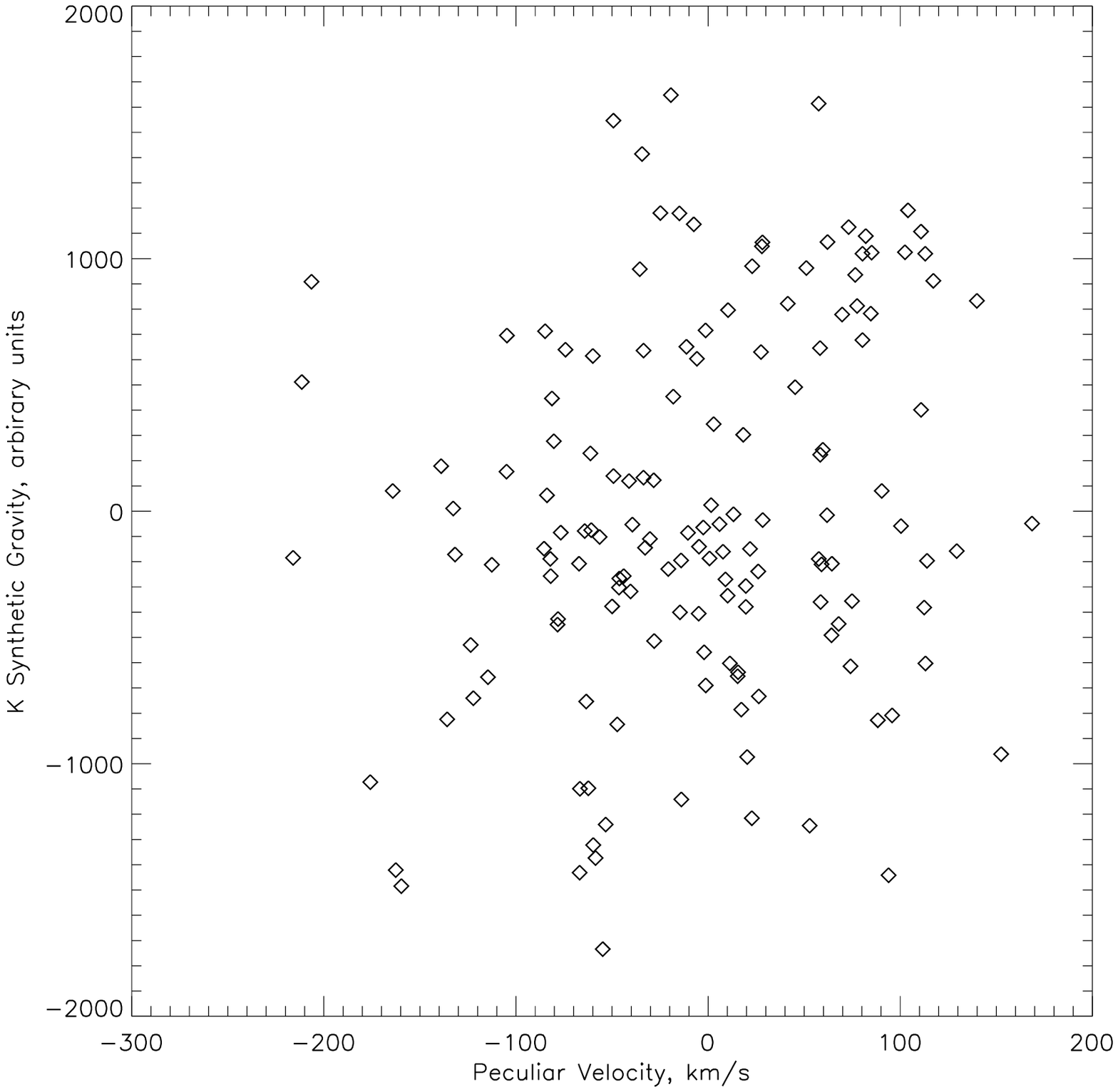}
\caption{Peculiar radial velocities of galaxies within the Local Volume (as
calculated from the isotropic-expansion kinematic model) plotted against the
radial component of gravitational force, assuming a constant ratio of mass to light in the
$B$ band (left) and $K$ band (right).}
\label{scatter1}
\end{figure}

The resulting plot for $B$ luminosity is shown in the left panel of Figure (\ref{scatter1}).
There is no apparent linear, quasi-linear
or frayed-S shape.  In fact there is no apparent correlation at all.  Postponing for a moment
seeking a correlation with more subtle statistical tools than the human eye, we will first
try to find a plot which gives something more obvious.

Perhaps the problem lies in using blue luminosities.  It is well known that starbursts can skew
the mass-to-light ratio in this band drastically; the nearby dwarf IC 10 is nearly as luminous as
the spiral M33 in $B$, but much fainter in the infrared.  Using $K$-band luminosity, which should
be a much better measure of the actual mass of the galaxies, we arrive at the
right panel of Figure (\ref{scatter1}).

There is, again, no visible correlation.  There are significantly fewer galaxies with $K$
photometry to contribute to the synthetic gravity field, 
but the effect of the unmeasured objects should be slight: none are as bright as the
Magellanic Clouds in $B$, and so the combined effect of all of them
is less than about that of one giant galaxy.

\begin{figure}
\plottwo{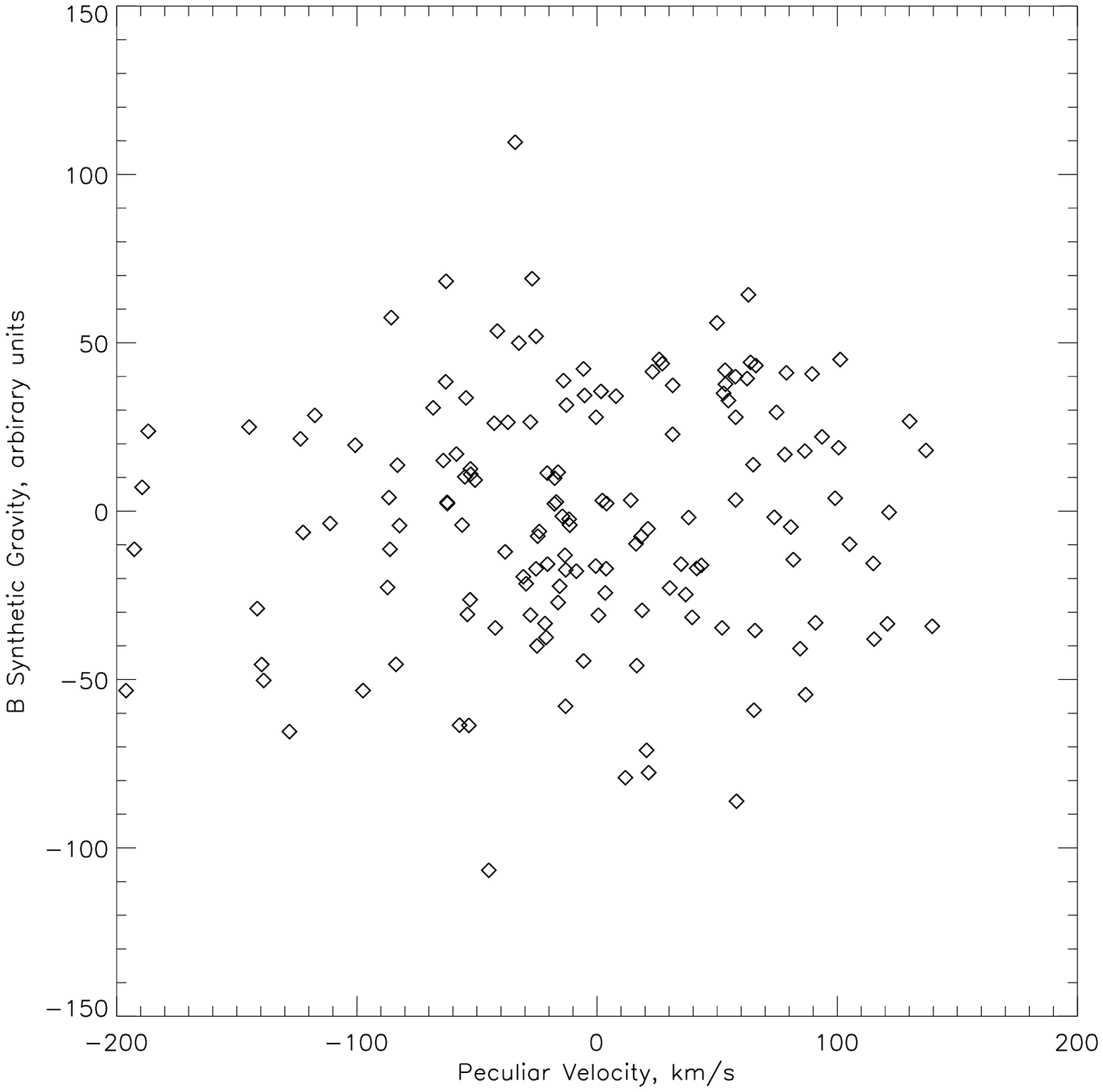}{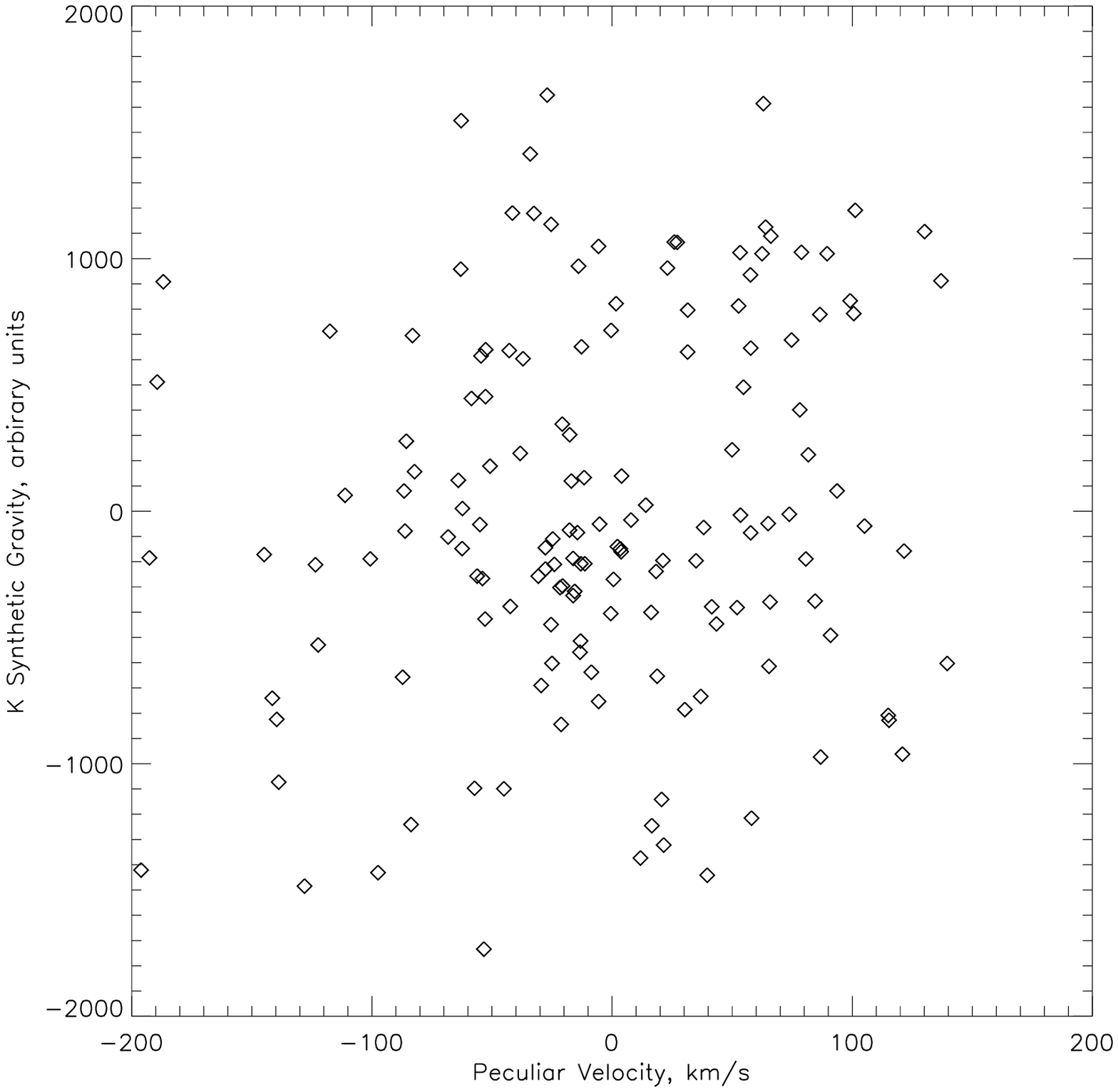}
\caption{Peculiar radial velocities of galaxies within the Local Volume plotted
against the reconstructed radial gravitational acceleration as before, but using
the tensor (anisotropic expansion) solution.  Again, $B$ band light is used on the left,
$K$ band on the right.}
\label{hscatter1}
\end{figure}

Recalling that the anisotropic model was calculated to be a better fit to the data, 
we turn to the peculiar velocities based on that model for Figure
(\ref{hscatter1}).  Again there is no improvement apparent.

\begin{figure}
\plottwo{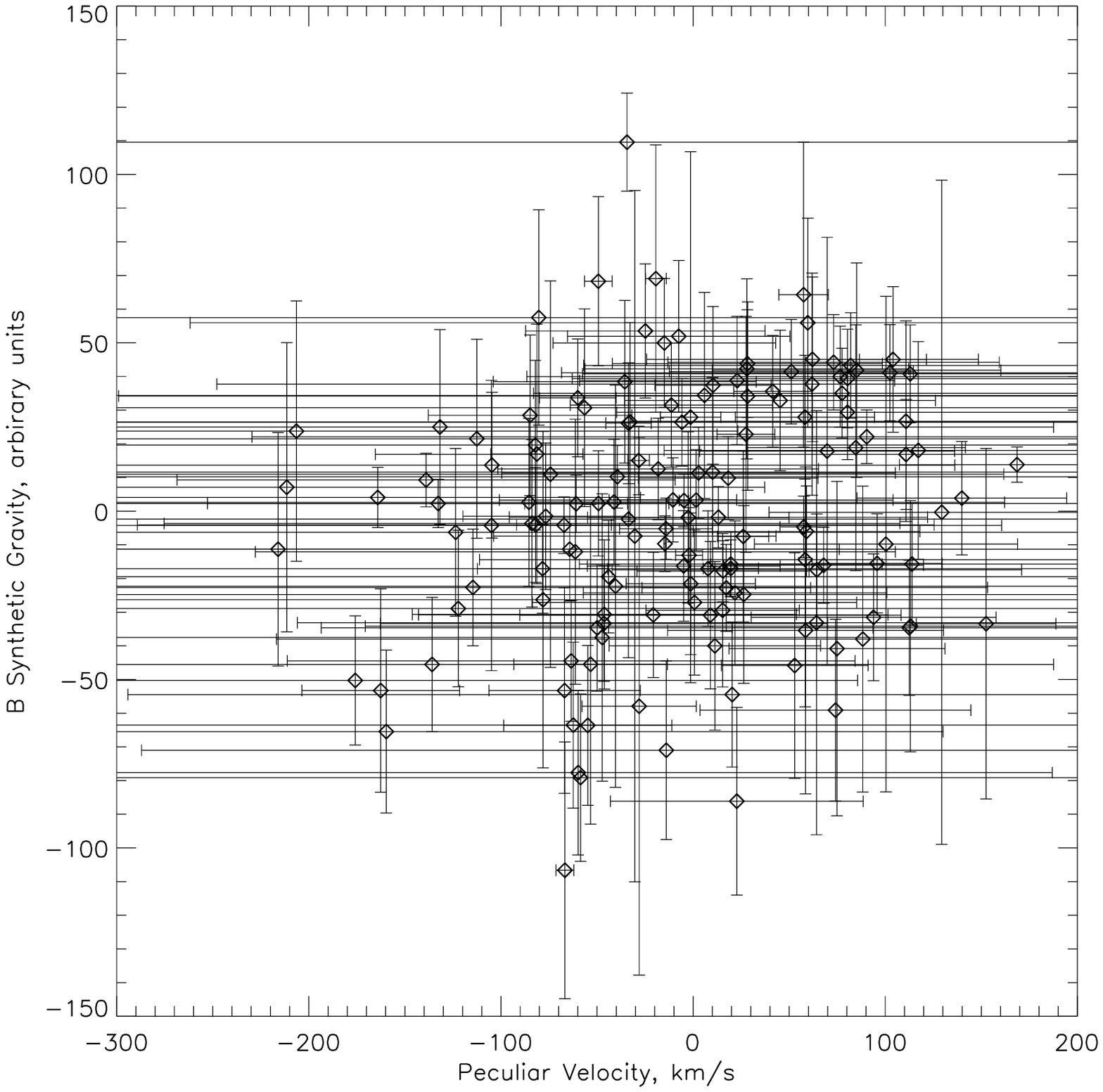}{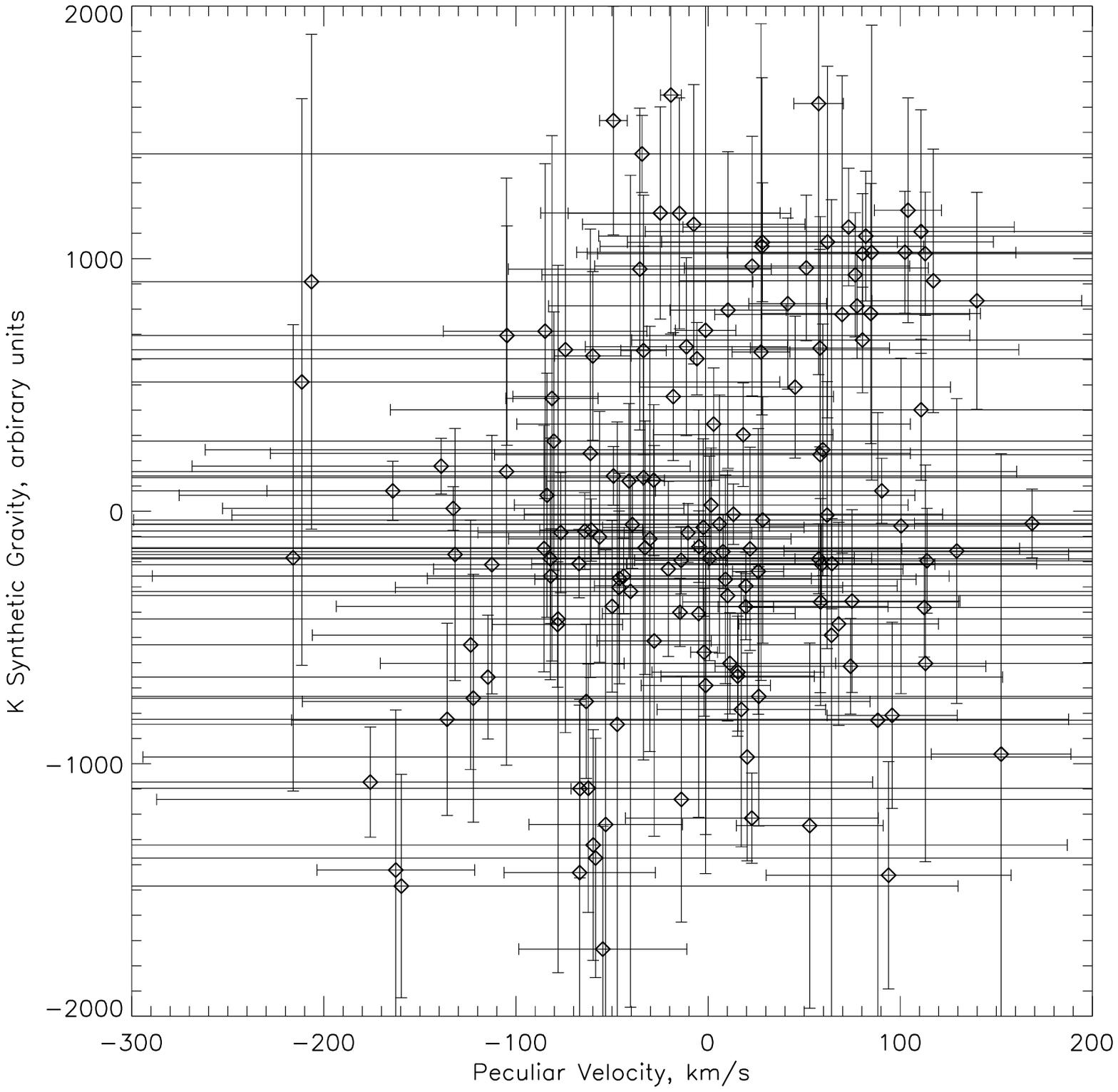}
\caption{As in Figure \ref{scatter1}, but with error bars included.  Uncertainties in
peculiar velocity are calculated from those in the model parameters and galaxy distances;
those in synthetic gravity, from uncertainties in distances and photometry.}
\label{scatter2}
\end{figure}

Perhaps the uncertainties in the various quantities are large enough to obscure any correlation;
that is, perhaps the data are simply not good enough yet to see a signal.
Adding the uncertainties in synthetic gravity and peculiar velocity
as error bars gives Figure (\ref{scatter2}).

\begin{figure}
\plottwo{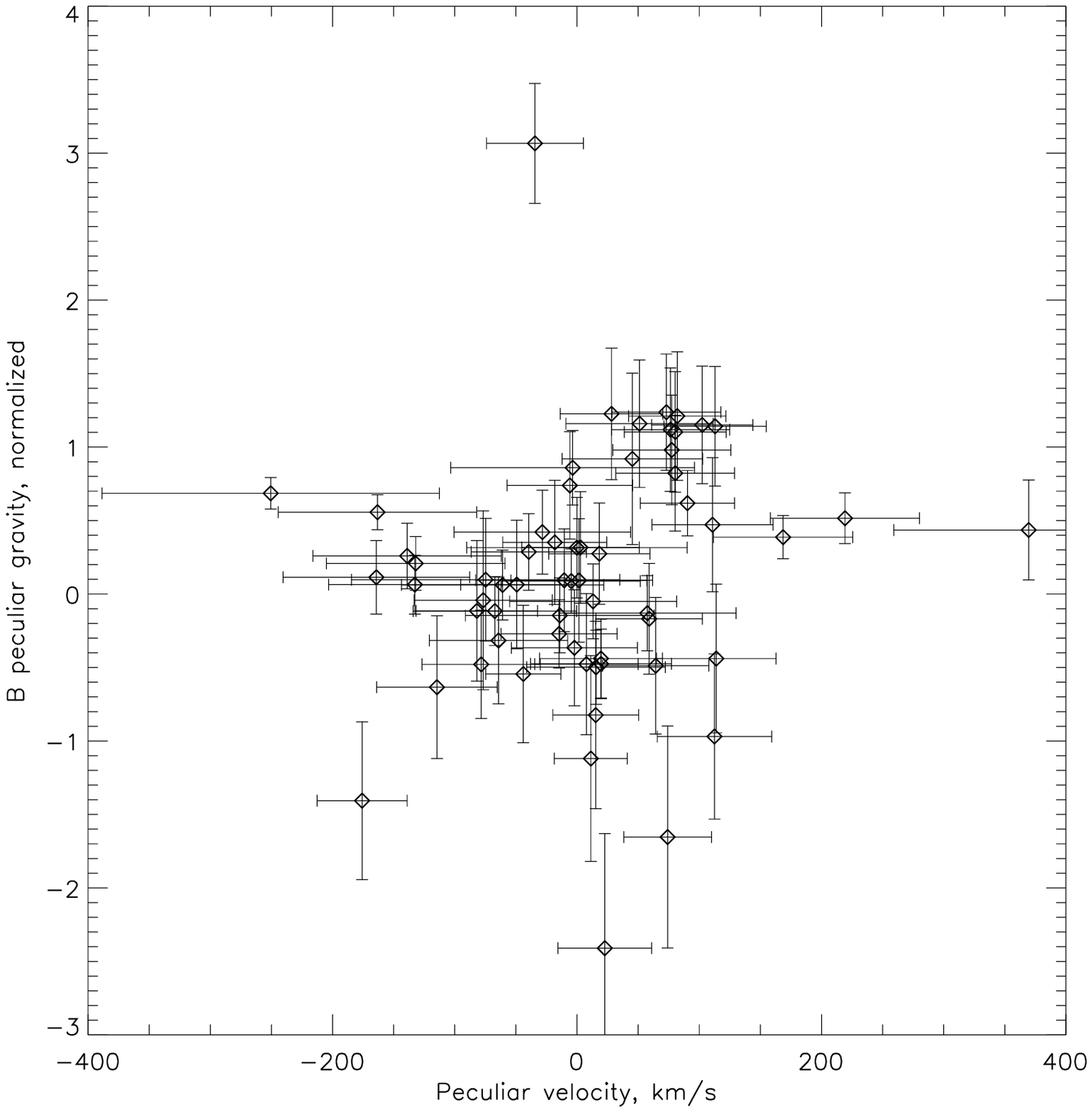}{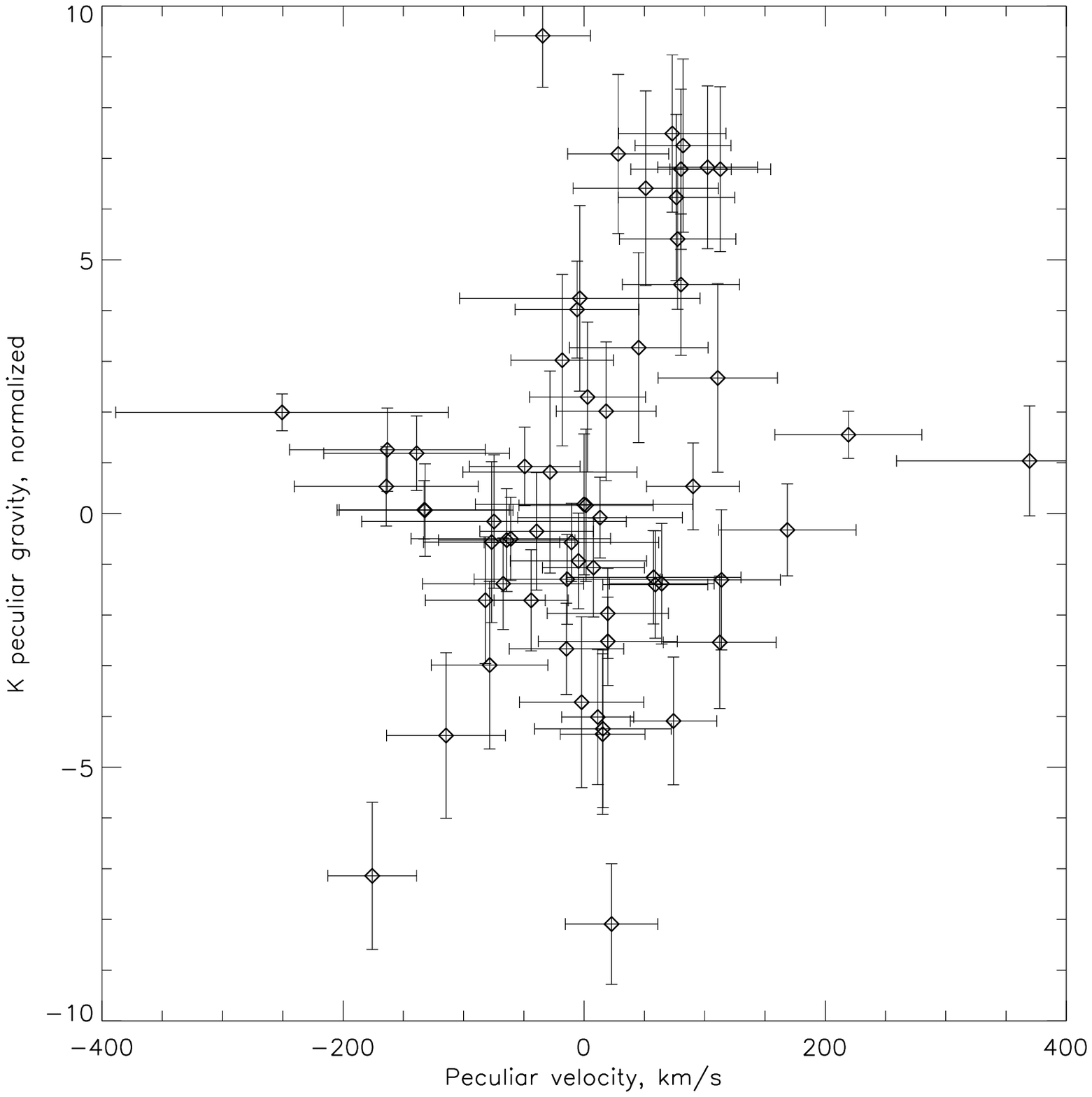}
\caption{As in Figure \ref{scatter2}, but including only the 63 points with the smallest
error bars.}
\label{scatter3}
\end{figure}

Some points on the plots are clearly very uncertain indeed.  Though it appears that
there are enough good data to delineate a trend, if it existed, including all the errors
serves to obscure the matter more than it helps.  Using only the 63 points with the best
synthetic gravity uncertainties, we have Figure (\ref{scatter3}).  As one would expect,
points have been
preferentially removed at synthetic gravities of large (positive and negative)
magnitude, since these are most subject to distance errors in nearby galaxies;
and there is a bit of structure due to small-number statistics.
However, no real trend is visible, and the errors here are certainly small enough to show one.

I have claimed that there is no correlation on any of the plots presented; however,
it is very difficult to prove the absence of {\em any} correlation.  (In fact
there could be a very strong correlation of a particular kind, say a high-frequency
sine wave, which would be entirely masked by the errors.  But we are looking for something
less general, a monotonic function.)  The fact that none is evident
in the plots here presented is a strong indication in that direction, the human eye being very good
at detecting correlations amidst noise (indeed, even when none exists); but we would like
something quantitative.

Consider a linear relation.  While we expect the Local Volume to be nonlinear as far as
gravitational effects are concerned, we should certainly be able to approximate it with
a straight line to some degree of accuracy.  If we fit one to the $K$-band data\footnote{From
here on calculations will be confined to the $K$ band peculiar gravities to save time
and space.  This band should show the effect of mass best, if any exists, and the
incompleteness of our data should not make any significant overall difference.} of Figure
(\ref{scatter1}), then subtract the fit, we are left with a peculiar velocity dispersion
of 77 km s$^{-1}$.  This is indistinguishable, statistically, from the raw dispersion; so
there is no linear fit.

In search of the most basic, qualitative signal, we then divide the plot into positive and
negative peculiar gravity halves, excluding any points whose errors would take them across
the zero line.  For each half we form the average and estimate the uncertainty in the average
based on the individual uncertainties in the peculiar velocities.
For the region of positive $K$ peculiar gravity, the average peculiar radial velocity is +27
km s$^{-1} \pm 7$; for negative, -23 km s$^{-1} \pm 6$.  The rms velocity dispersion for
both positive and negative K is 49 km s$^{-1}$.  But again, the uncertainties in peculiar velocity
are correlated, so the stated errors could be very misleading; and the dispersions are
much larger than the averages in magnitude.  There could be a general, average correlation
of peculiar velocity with synthetic gravity, but at best it explains a minority of the actual
motion, and it is not certain it exists.

It is time to examine the model itself in terms of its dynamical implications.

\subsection{A Dynamical Background Model}

Up to this time we have used the average, kinematically-derived background model as a
basis for peculiar velocities, and have found no correlation between velocities
(apples) and luminosity-derived peculiar gravity (oranges).  Suppose we try a
comparison which adjusts the model to minimize the
difference between the peculiar velocity and the $K$-band synthetic gravity:
\begin{equation}
\sigma^2 = \frac{1}{N} \sum_i \left(g_i - v_{pi} \right)^2 =
\frac{1}{N} \sum_i \left( g_i - (v_{\rm obs} - \left( {\bf v}_0 \cdot
\hat{\bf r} + \hat{\bf r} \cdot {\bf H} \cdot {\bf r} \right) \right))^2
\label{eq:dispersion2}
\end{equation}
We are faced with the fact that the normalization of the synthetic gravity, $g_i$, is
unknown.  If it is too small, we are essentially reproducing the kinematic model;
if it is too big, we are fitting a model to the luminosity field and ignoring
motions---a calculation which might be of some interest, but not to us now.

\begin{figure}
\plottwo{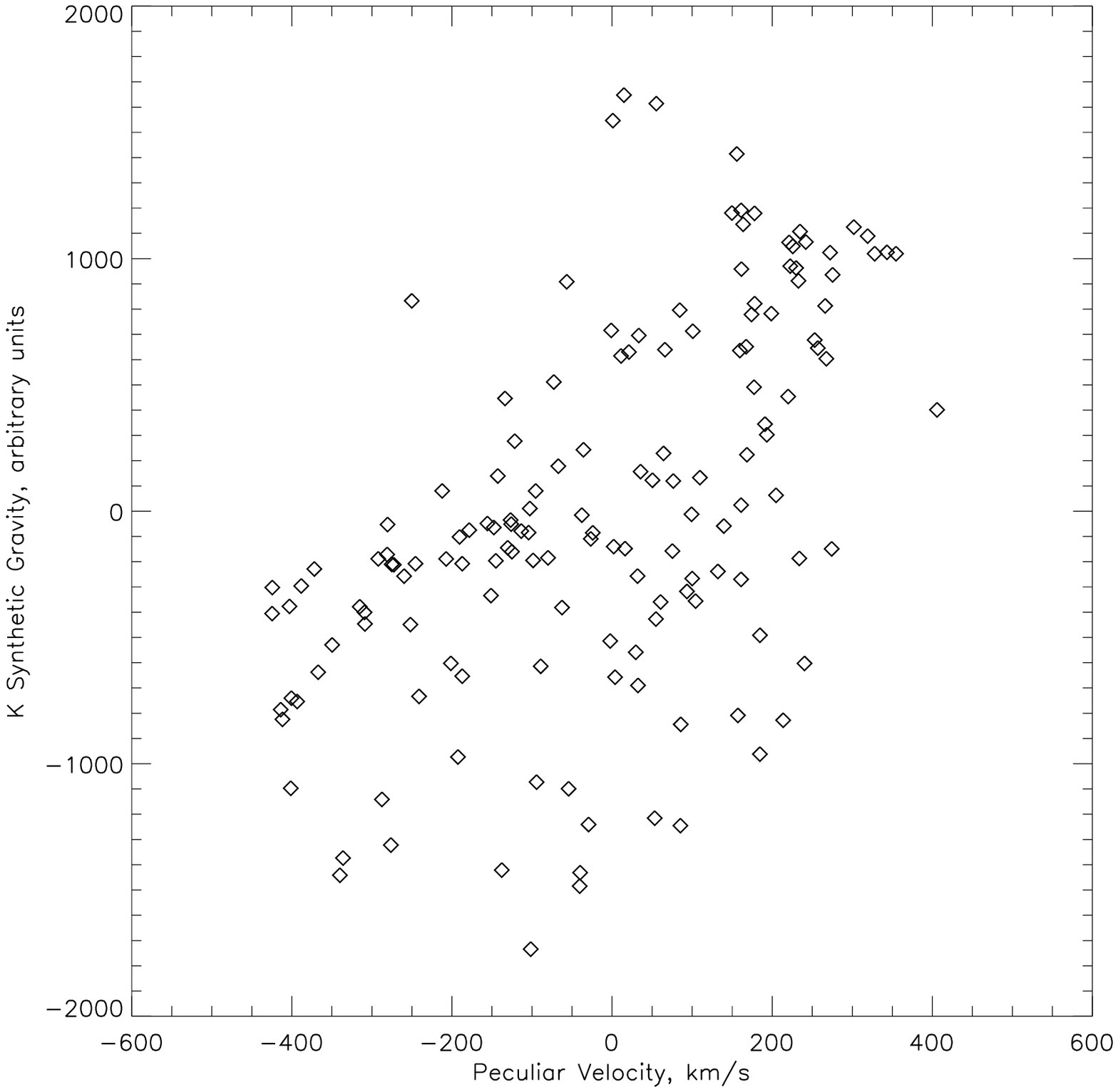}{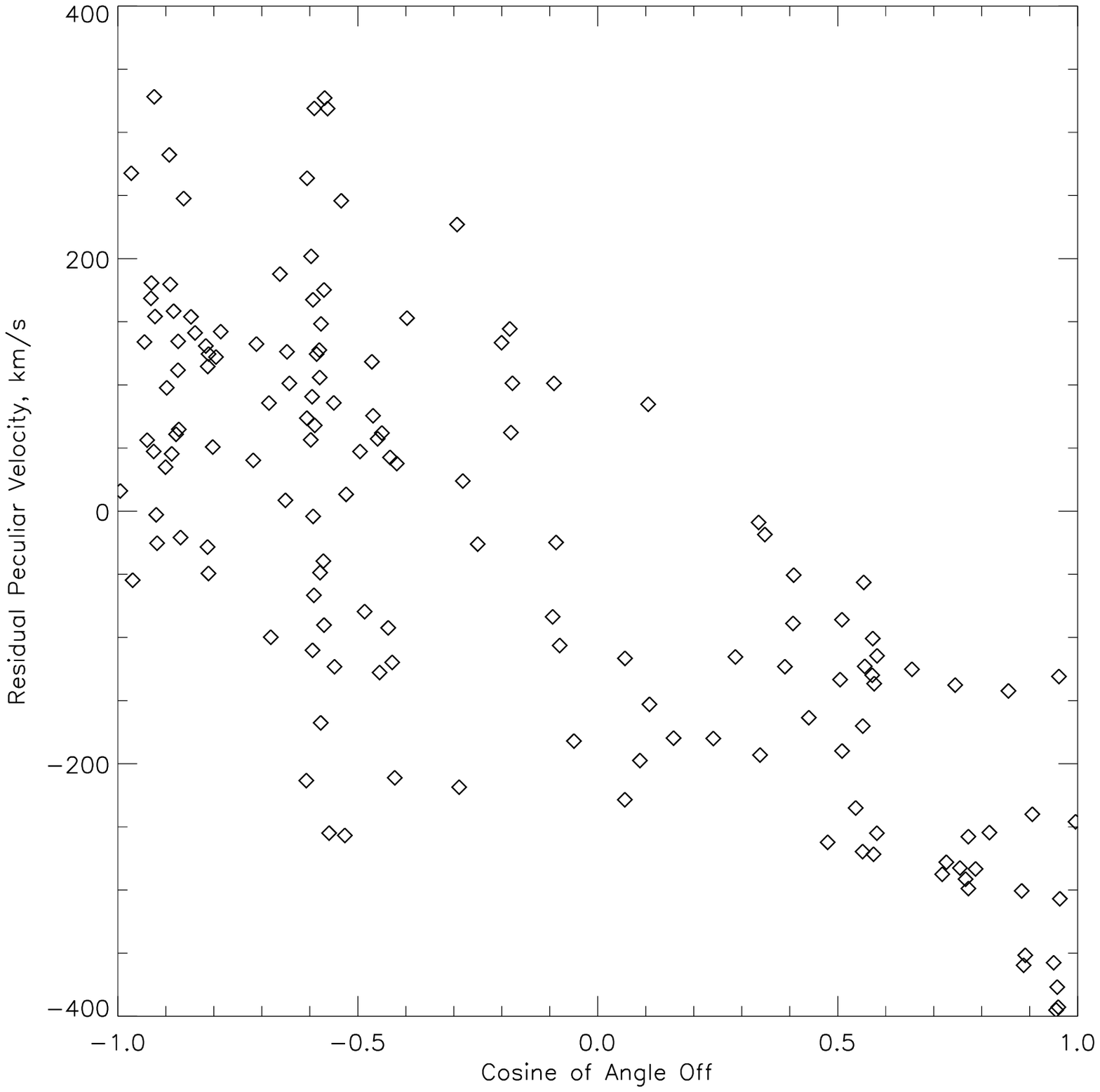}
\caption{Left, the radial component of gravity, assuming mass is proportional to $K$-band
light, smoothed on 1.2 Mpc scales, plotted against peculiar velocity relative to
a model adjusted to give the best correlation.  Right, the residuals of the left-hand
plot after a linear fit is subtracted, plotted against the cosine of the angle between
the particular galaxy and the difference in constant vectors between the kinematic
and dynamic solutions.  The clear trend in the right-hand plot shows the trend in
the left-hand plot to be spurious; see text.}
\label{dynamic}
\end{figure}

In practice a series of models was calculated with different normalizations, starting
with one which matched the rms value of the peculiar velocity in the isotropic
kinematic model.  This turned out in fact to give the best correlation, shown in 
the left-hand panel of
Figure (\ref{dynamic}), with parameters shown in Table (\ref{solutions2}).

Finally we have an undoubted correlation.  There is significant scatter, but it is
no longer possible to say that there is {\em no} relation between luminosity and
peculiar velocity.  (An anisotropic solution, calculated at the same time, gave
an essentially identical scatter.)

How good is the correlation?  If we fit a line, as we did for the $K$-pv plot
in the kinematic model, then subtract it, we arrive at a velocity dispersion
of 175 km s$^{-1}$.  This is significantly better than the 209 km s$^{-1}$ of
the uncorrected model;
the F-ratio test gives a 98\% chance of the smaller dispersion being drawn from
a different distribution.  

It is still much larger than the dispersion around
the best-fit kinematic model.  This is worrying; why should a model showing more
insight into the physics give a worse overall correlation?  Not necessarily troubling,
but adding nothing to the situation, is the fact that neither the new ${\bf v}_0$
nor the difference (the vector added to the kinematic solution to get the new one)
points in the direction of anything interesting outside the Local Volume.  It
would be interesting, say, if the new vector pointed toward Virgo, or the
difference vector in the direction of the CMB dipole.
Still, surely one cannot find a correlation which isn't there?

\begin{deluxetable}{lcrr}
\tablecaption{The Dynamical Background Solution \label{solutions2}}
\tablehead{
\colhead{Quantity} & \colhead{Magnitude} & \colhead{L} & \colhead{B} }
\startdata
 & km s$^{-1}$ or km s$^{-1}$ Mpc$^{-1}$ & Degrees &Degrees  \\
Dynamical Model & & & \\
V & 414 & 223 & -20  \\
H & 91& & \\
Dynamical minus Kinematic & & & \\
V & 325 & 265 & 13 \\
H & 25 & & \\
\enddata
\tablecomments{Parameters of the dynamical background model (constant vector
plus uniform expansion, maximizing the correlation between peculiar velocity
and $K$-band luminosity).}
\end{deluxetable}
\clearpage

One can.  Consider a Hubble plot of a galaxy group (radial velocity versus distance).
Suppose there is no correlation of peculiar velocity with luminosity, so that
there is a constant dispersion about the average line.  Synthetic gravities will
be positive on the near side and negative on the far side, with no evident relation
with peculiar velocity.

Next suppose that, instead of the ``proper'' model, one imposed a higher Hubble
constant (steeper slope); now the group as a whole will show negative peculiar
velocities.  Add a constant vector in the direction away from the group, and all their
peculiar velocities will increase.  If they increase to the point that the new model fit passes 
near the center of the group the average peculiar velocity will again be zero,
but now there will be more points above the line (positive peculiar velocity)
on the near side and more below the line (negative peculiar velocity) on the far
side; these will correlate with the synthetic gravity.

If groups were distributed evenly around the sky the better correlation thus gained
for one would be offset by a worse correlation for another on the far side.  But
the distribution of groups in the Local Volume
is not even or random, so this could be the explanation
for the correlation seen in the left-hand side of Figure (\ref{dynamic}).  There
would be three clear signs of this effect: (i) the added constant vector would be in the
direction away from most galaxies; (ii) there would be a greater dispersion
away from the correlation to negative peculiar velocities (whole groups brought
below the Hubble line) than positive; and (iii) there will be a correlation
between the residuals (after the spurious correlation is subtracted from
peculiar velocities) and the direction of the added constant vector.

Condition (i) is satisfied; the added constant vector is less than $23\arcdeg$
away from the antipode of
the average of all radius vectors of Local Volume galaxies in our sample
(this has less than 5\% chance of happening at random).  
Condition (ii) is evident from the left-hand side of Figure(\ref{dynamic}),
and condition (iii) from the right-hand side.  Our ``correlation'' found using
the dynamic criterion is spurious.  
Perhaps more important than this specific result, the sensitive dependence of peculiar velocities on the
background model is highlighted.  

\section{Discussion and Conclusions}

\subsection{Other Results in the Local Volume}

The main result of this study, that there is no apparent relation between
luminosity and peculiar velocity on megaparsec scales in the Local Volume,
is at odds with some previous work.  Signs of the expected gravitational
effects of galaxy groups have been seen and masses have even been calculated. 
The disagreement is mostly due to the fact that two different questions have
been asked.  Other work has asked, ``What is the mass of this galaxy group?''
while I have asked, ``Is there a detectable mass?''  It should not be
surprising that different questions come up with different answers.

Additionally, the significance of the fact that one sees an expected sign of the effect of
luminous mass on peculiar velocity depends greatly on the context.  It is entirely
fair and correct to note in passing that a feature on a plot has a ready explanation
under plausible assumptions.  However, if one is specifically looking for these
signs and finds some of them but not others, the positive results are much
less likely actually to mean that the assumptions are confirmed.  Especially given
the sensitivity of peculiar velocity studies to the background model as shown above,
the standard of proof must be set higher than the occasional positive result.
If it looks like a duck, but doesn't quack quite like a duck or fly at all like a duck, it
probably isn't a duck.  (It may be a decoy.)  It is certainly possible that,
for instance, gravitating mass might be found near a visible galaxy group now
and then by chance.

Note that {\em we do not know} the background against which peculiar velocity studies
should take place.  We have an average Hubble constant and solar reflex velocity
computed from the data; these are statistics, which we hope approximate the real
background well enough.  Similarly, the uncertainties in background model parameters
(and hence peculiar velocities) are based on how well the model fits the data at hand;
they are not absolute.

That said, there have been previous indications that galaxy groups in the Local Volume
do not have the expected gravitational effect on peculiar velocities.  \citet{KSD03}
found the Canes Venatici Group's kinematics to be mostly a matter of Hubble
expansion, as did \citet{JFB98} for the Sculptor Group (the latter result has
been supported by other studies).  \citet{WH03} found that there
was no significant difference in velocity dispersion within the Local Volume between galaxies
within groups and field galaxies.

\subsection{Results on Other Scales}

The main result of this study also appears to be at odds with some results on smaller and
larger scales.  The venerable timing argument
(whether in the original two-body form of \citet{KW59} or with dwarf galaxies
added, as in \citet{WH99} for instance) gives reasonable results for the mass
and age of the Local Group; the more sophisticated Least-Action approach of
\citet{P90} (among other implementations) also seems to work well.  On larger
scales, as has been noted, there is agreement between the luminosity and
inferred mass fields; in addition, a version of the Least Action reaching down
to Local Volume scales has been successful here \citep{SJ01}.  While I
have been careful to note that the present work does not directly apply to
other scales, it is worth
looking at how much of a disagreement there is and how it might be resolved.

There is the possibility of a fortuitous result, for instance that a mass
of dark matter is located by chance in such a way as to produce motion
consistent with light tracing mass in one area, while such a condition is
not general in the Volume.  A good result with the Local Group then
becomes a matter of small-number (one) statistics.

Small-number statistics can also enter in when checking how good a result is.
Both \citet{P90} and \citet{SJ01} use a few galaxies (six and twelve, respectively)
to test their fits.  For comparison, \citet{WH03} showed that the fourteen
galaxies used by \citet{EB01} to characterize Local Volume motions gave a quite
misleading picture.  The same thing could be happening here (and note that one
of the galaxies in \citet{SJ01} did not fit well at all).

But mostly it appears that there are differences in interpretation as to
what constitutes a good result.  The timing argument
produces a number in which the age and mass of the Local Group are mixed.
Given an age in accordance with other determinations, one finds a mass which is
in reasonable agreement with other determinations.  But the latter is not
really known independently to within a factor of two (or more), so 
the timing argument produces am age consistent
with other results only within this factor.  Moreover, as the plot in \citet{WH99}
shows, there is a large uncertainty (certainly another factor of two)
in the combined age-mass number.  The best that can be said is that the timing
argument shows that its assumptions produce a result which is roughly consistent
with other determinations.

The calculations of \citet{P90} and \citet{SJ01} produce much tighter correlations
between calculated and observed redshifts for their test galaxies.  But if
one examines instead the correlation between predicted and observed peculiar
velocities, the correlation looks much looser, to the point where it might not
be significant in the context of a 75 km s$^{-1}$ rms dipersion.  Probably
the best interpretation is that
these studies have arrived at the right overall density of mass, as well as
a good approximation of fluctuation amplitude; but (in light of the present
work) may not have located those fluctuations correctly.

\subsection{Conclusions and Implications}

Given the overall result that the peculiar velocity field and the luminosity field
do not correlate, one of two conclusions follow:
\begin{itemize}
\item Either light does not trace mass in the Local Volume, or
\item Peculiar velocities are not produced by gravitational interaction.
\end{itemize}
Just how light might fail to trace mass has not been shown in any detail.  There could
be a large population of ``dark galaxies'', such as those postulated by \citet{JHH97}
on the basis of lensed quasars.  Attempts to reconstruct the local dark matter field
from the peculiar velocity field using a POTENT-type algorithm (see, for example, \citet{BDF90})
have not been successful.  It appears that 149 data points are just too few (POTENT calculations
use galaxies in the thousands).
It does appear that theories which dispense with dark matter by modifying the behavior
of gravity are ruled out by this result.

It is even less obvious how peculiar velocities might be produced if gravitational
interaction is not the source.  The only competing explanation, primeval turbulence,
seems to have been ruled out long ago (see \citet{P93}, pp. 541ff).

This result does {\em not} mean that mass and light are unrelated on larger scales
(indeed, as noted in the Introduction, they seem to go together well there). 
Nor does it mean that they are unrelated on galaxy scales, where clearly dark matter
haloes are concentric with luminous matter.  Nor does it mean that they are unrelated
on Mpc scales in galaxy clusters, where again they seem to be concentric.  It
{\em does} mean that a plausible and convenient assumption commonly made must be
abandoned; but in return we have gained some important clues as to the behavior
(and, one hopes eventually, the nature) of this elusive dark matter.

\acknowledgements

The author gratefully acknowledges help from discussions with many collegues, especially
Dr. Donald Lynden-Bell, Dr. Nicholas B. Suntzeff and Dr. David Spergel; and would
like to express his sincere appreciation for all the observers whose data appear in
Table (\ref{data1}).

This research has made use of the NASA/IPAC Extragalactic Database (NED) which is
operated by the Jet Propulsion Laboratory, California Institute of Technology,
under contract with the National Aeronautics and Space Administration.

\end{document}